\documentclass[fleqn,usenatbib]{mnras}
\usepackage{newtxtext,newtxmath}

\usepackage[T1]{fontenc}

\DeclareRobustCommand{\VAN}[3]{#2}
\let\VANthebibliography\thebibliography
\def\thebibliography{\DeclareRobustCommand{\VAN}[3]{##3}\VANthebibliography}

\usepackage{graphicx}	
\usepackage{amsmath}	

\title[Jet precession in the blazar AO 0235+164]{Parsec-scale jet precession and a putative supermassive binary black hole system in the blazar AO 0235+164}

\author[F. B. da Silva Junior and A. Caproni]{
F. B. da Silva Junior,$^{1}$\thanks{E-mail: flavio.junior94@cs.unicid.edu.br (KTS)}
and A. Caproni,$^{1}$
\\
$^{1}$Núcleo de Astrofísica, Universidade Cidade de São Paulo, R. Galvão Bueno 868, Liberdade, São Paulo, SP 01506-000, Brazil\\
}

\date{Accepted 2025 July 17. Received 2025 June 23; in original form 2024 December 26}

\pubyear{2025}

\begin{document}
\label{firstpage}
\pagerange{\pageref{firstpage}--\pageref{lastpage}}
\maketitle

\begin{abstract}
The blazar AO\,0235+164 is a key source for studying the interplay between multi-wavelength variability in its light curves and changes in the position angles and apparent velocities of its parsec-scale jet components. In this work, we analyse public interferometric radio maps of AO\,0235+164 at 15 and 43 GHz, using the Cross Entropy global optimisation technique to determine the structural parameters of its jet components. We identified 36 kinematically distinct jet components across all sky quadrants, indicating a highly relativistic parsec-scale jet with a minimum Lorentz factor of $34\pm7$ and a maximum viewing angle of $37\degr\pm8\degr$. The temporal evolution of these jet components was modelled as a relativistic jet under a constant precession rate. The optimal clockwise precession model has a precession period of $8.4\pm0.2$  years, consistent with the 8.13-year periodicity previously detected in optical light curves, besides providing a time-variable Doppler boosting factor correlated with the most intense flares at gamma-ray energies. For the counter-clockwise precession, a period of $6.0\pm0.1$ years is found, compatible with the 5–6-year periodicities detected at radio and optical wavelengths. It is plausible that a supermassive black hole binary system in the nucleus of AO\,0235+164 drives the parsec-scale jet precession and induces nodding motions consistent with short-term continuum periodicities. Nonetheless, alternative scenarios (e.g., intrinsic curved jet, warped accretion disc instabilities, Lense-Thirring/Bardeen-Petterson effects, dual jets) cannot be ruled out as causes or optional explanations for the precession.

\end{abstract}




\section{Introduction}

Blazars are one of the most extreme types of active galactic nuclei (AGNs)
, with their jets forming a small angle relative to our line of sight (e.g., \citealt{1984RvMP...56..255B, 2019ARA&A..57..467B}). It is commonly believed that such objects have a supermassive black hole (SMBH) surrounded by an accretion disk and/or optically thick plasma, a dusty torus, and a jet composed of multiple components usually moving at superluminal speeds (e.g., \citealt{1993ARA&A..31..473A, Urry_Padoavani_1995, 2015ARA&A..53..365N}). Besides, these objects typically exhibit strong variability in their non-thermal continuum from radio to gamma-rays (e.g., \citealt{Urry_Padoavani_1995, 1997ARA&A..35..445U}).

Located at a redshift of 0.94 \citep{1987ApJ...318..577C}, the blazar AO\,0235+164 resides in a galaxy-rich field, potentially affected by gravitational lensing effects \citep[e.g.][]{1988A&A...198L..13S, 1993ApJ...415..101A, 2000AJ....120...41W}. AO\,0235+164 exhibits strong variability across the entire electromagnetic spectrum \citep[e.g.][]{1988LNP...307..176W, 2008A&A...480..339R, 2009ApJ...696.2170R, 2010ApJ...710.1271A}. Quasi-periodic radio variability has been reported in the literature, with periodicities around 1.8, 2.8, 3.7, 5.7, 10.0, 12.0 years \citep{2000ApJ...545..758R, 2001A&A...377..396R, 2006ApJ...650..749L, 2007A&A...462..547F, 2016Galax...4...17F, 2021MNRAS.501.5997T}. In the optical regime, periodicities of approximately 0.5 1.3, 1.5, 2.8, 3.6, 5.8 and 8.2 years were also detected in the case of AO\,0235+164 \citep[e.g.][]{1988LNP...307..176W, 1995PASP..107..863S, 2001A&A...377..396R, 2002A&A...381....1F, 2006A&A...459..731R, 2008A&A...480..339R, 2014Ap&SS.351..281W, galaxies4030017, 2017ApJ...837...45F, 2022MNRAS.513.5238R, 2023MNRAS.518.5788O}. Short-term variability, particularly at lower frequencies, may be caused by interstellar scintillation of ultra-compact components with a size of approximately 10 $\mu$as \citep[e.g.][]{2006ApJS..165..439R}.

Radio images of AO\,0235+164 at kiloparsec scales show a "one-sided halo" morphology \citep[e.g.][]{2007ApJS..171..376C}, characterized by an extended structure on one side of an angularly unresolved nucleus that is surrounded by diffuse halo-like emission. AO\,0235+164 is partially resolved in most radio bands when observed using Very Long Baseline Interferometry (VLBI) techniques. Previous VLBI studies of this source found a jet-like morphology at parsec scales\citep[e.g.][]{1984ApJ...284...60J, 1996A&A...307...15C, 1999NewAR..43..707C, Jorstad_et_al_2001, 2018MNRAS.475.4994K}. 

Association of flaring episodes with the emergence of superluminal jet components in AO\,0235+164 was also reported in the literature \citep[e.g.][]{2011ApJ...735L..10A, 2024A&A...689A..56E}. The observed brightness temperature of the core region of AO\,0235+164 is exceptionally high, reaching values as high as $10^{13.8}$ K \citep{2000PASJ...52..975F}, which exceeds the upper limit set by the Inverse Compton process (e.g., \citealt{1969ApJ...155L..71K}), even after considering extremely high Doppler boosting factors. 

In this work, the study of public interferometric radio maps of AO\,0235+164 obtained along almost the last 30 years has allowed the kinematic identification of 36 parsec-scale jet components moving relativistically from the core.

We show that changes in their position angles and apparent velocities over time are compatible with the precession of a relativistic jet with a period of approximately 8.4 years (or 6.0 years in the case of a counter-clockwise sense of precession). The feasibility of this jet precession being caused by a supermassive black hole binary system (SMBHB) is analysed, assuming that the orbital plane of the secondary supermassive black hole does not coincide with the primary accretion disc, which induces torques that cause the precession of the disc and the jet \citep[e.g.][]{2000A&A...355..915A, 2000A&A...360...57R, 2004MNRAS.349.1218C, 2004ApJ...602..625C, 2006ApJ...653..112C, 2013MNRAS.428..280C}.

This work is structured as follows: In \autoref{sec:ObsData}, it is provided a description of the interferometric radio data used in this work. The process of kinematic identification of the jet components in AO\,0235+164, the estimates of lower limit for the jet bulk Lorentz factor, the upper limit for jet viewing angle, and the determination of the brightness temperature of the core region are presented in \autoref{sec03}. Temporal changes in the apparent velocity and position angle of the jet components are discussed in the context of the jet precession scenario in \autoref{sec:Disc}. From the best of our knowledge, the variability of the brightness temperature of the core region is also included in the precession modellings for the first time in the literature. In \autoref{sec04}, we show that this jet precession may be related to the existence of a supermassive binary black hole system in AO\,0235+164. The same putative binary scenario can also driven nodding motions, which are able to introduce additional (periodic) changes in the jet viewing angle, impacting the observed jet velocity and the jet position angle. Some of the short-term periodicities reported in the literature are also discussed in terms of this binary system in \autoref{sec04}. Alternative scenarios for jet precession or different mechanisms for driving jet precession in AO\,0235+164 are presented in \autoref{AlterScenarios}. Final remarks are made in \autoref{sec05}.

Throughout this work, the $\Lambda$CDM cosmology is adopted with $H_0=71.0$ km s$^{-1}$, $\Omega_\mathrm{M} = 0.27$, $\Omega_\mathrm{\Lambda} = 0.73$, which, for the redshift of AO\,0235+164, leads to a luminosity distance $D_\mathrm{L}= 6141.7$ Mpc, and a linear scale of 7.91 pc mas$^{-1}$. It establishes a relationship between proper motion and linear velocity, resulting in 1.0 mas year$^{-1}$ = 50.07$c$ for AO\,0235+164, where $c$ is the
speed of light.

\begin{figure}
	
	\includegraphics[width=1.0\columnwidth]{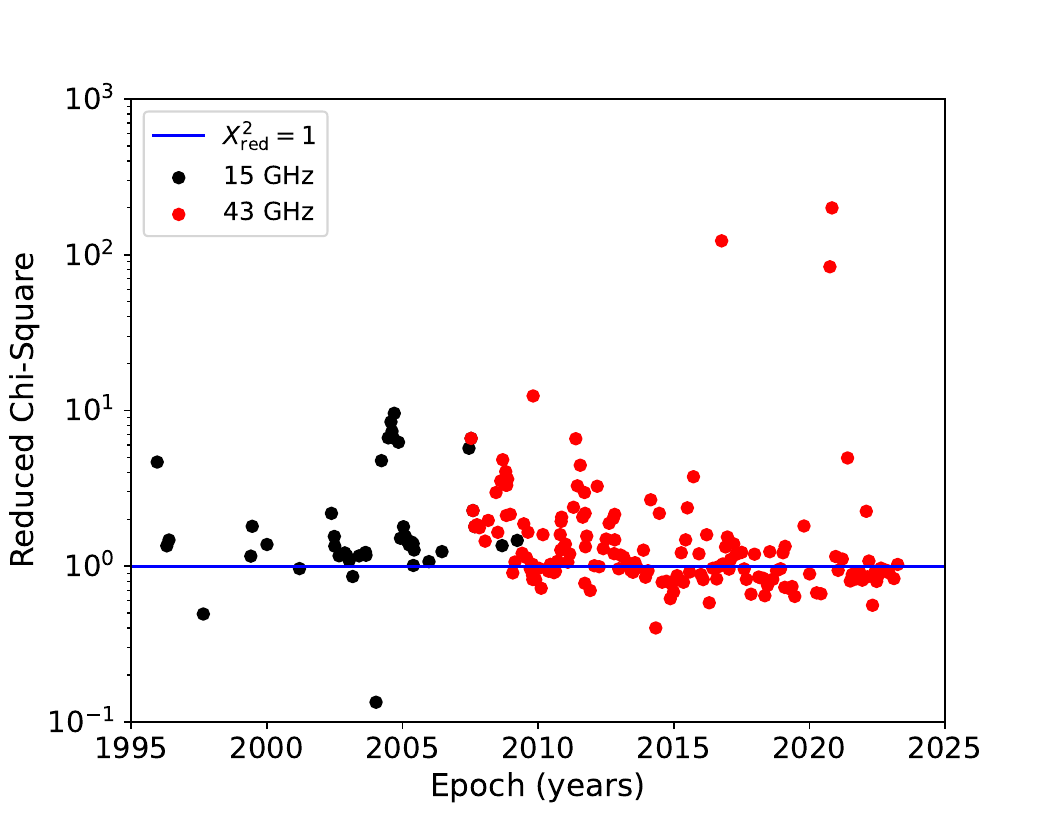}
        \centering
    \caption{Reduced chi-squared values derived by fitting the complex visibility data of AO 0235+164 at 15 GHz (black circles) and 43 GHz (red circles) with elliptical Gaussian components, whose structural parameters were determined in the image plane via the CE optimisation technique. The blue horizontal line marks the unitary value for the reduced chi-square. 
    }
   \label{Chi-Square} 
\end{figure}

\begin{figure*}
	\includegraphics[width=1.0\textwidth]{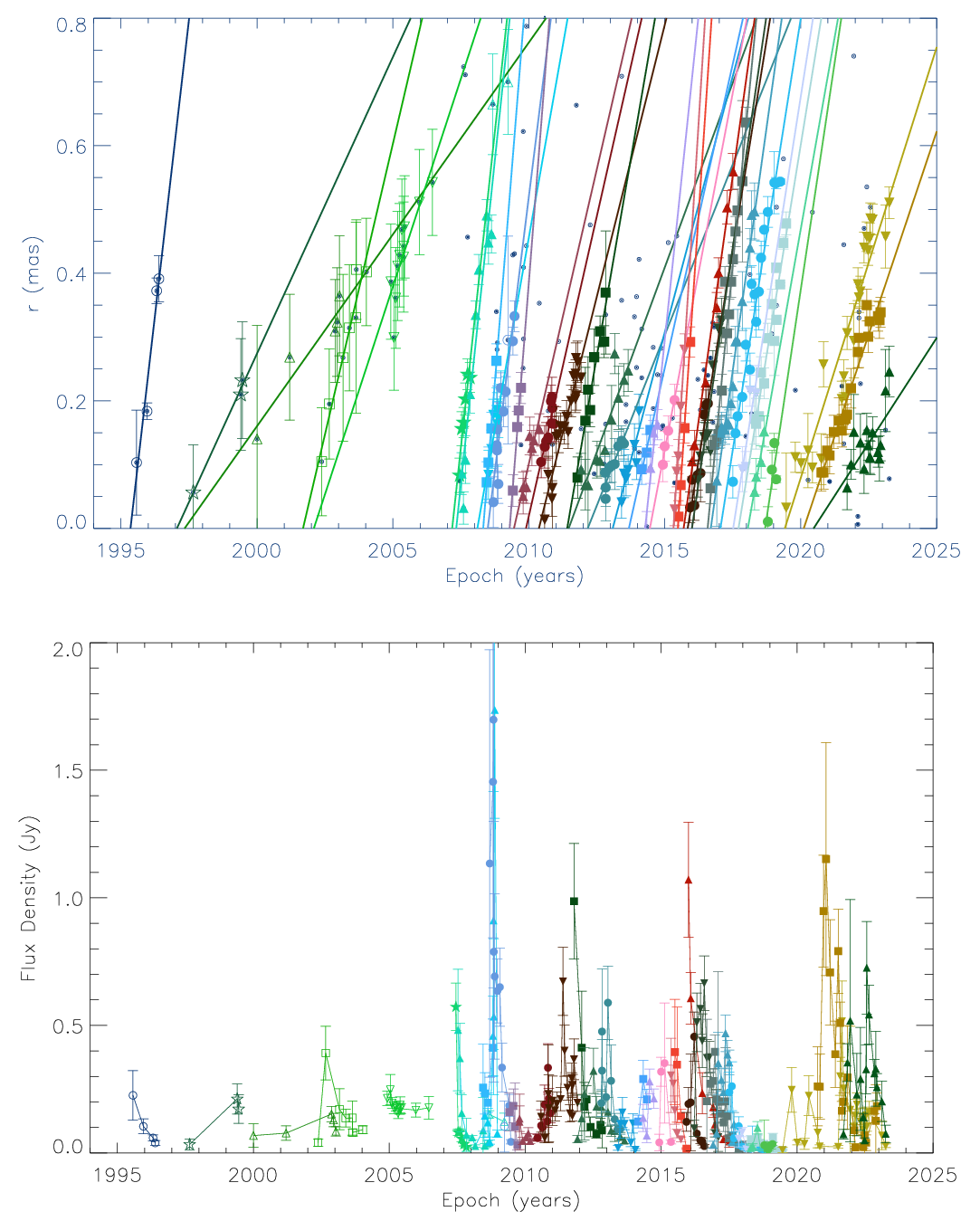}
        \centering
    \caption{{\it Top panel}: Time evolution of the core-component distance of the 36 jet components of AO\,0235+164 identified in this work. Open symbols represent jet components identified at 15 GHz, while filled symbols correspond to those at 43 GHz. Straight lines represent the linear regression for each individual component, while gray points are unidentified jet components from the kinematic point of view. {\it Bottom panel}: Flux density behaviour of the same 36 jet components.}
   \label{rxt_Fluxxt}
\end{figure*}

\begin{table*}
	\centering
	\parbox{12.5cm}{\caption{Kinematic parameters of the CE model-fitting jet components of AO\,0235+164 identified in this work.} 	\label{table:Kin_Param_AO0235_table}} 
	\begin{tabular}{ccccccc}
		\hline
  Jet component &  $t_{0}$ & $\mu$ & $\beta_{\mathrm{app}}$ & $\bar{\eta}$ & $N_{\mathrm{epoch}}$ $^\mathrm{a}$ & $\bf \textit{r}_{\mathrm{\bf p}}$ \\ & (yr) & ($\mu$as\,yr$^{-1}$) & & (deg) & \\
		\hline 
C1	&	1995.35	$\pm$	0.12	&	370.8	$\pm$	57.3	&	18.6	$\pm$	2.9	&	-20.6	$\pm$	7.6	&	4 &  0.980 \\
C2	&	1997.06	$\pm$	0.54	&	93.0	$\pm$	25.6	&	4.7	$\pm$	1.3	&	-37.6	$\pm$	3.8	&	3 &  0.996 \\
C3	&	1997.34	$\pm$	1.10	&	60.2	$\pm$	13.7	&	3.0	$\pm$	0.7	&	-73.2	$\pm$	4.1	&	5 &  0.949 \\
C4	&	2001.69	$\pm$	0.24	&	182.4	$\pm$	26.0	&	9.1	$\pm$	1.3	&	-77.3	$\pm$	2.9	&	7 &  0.970 \\
C5	&	2002.09	$\pm$	0.68	&	130.3	$\pm$	26.5	&	6.5	$\pm$	1.3	&	-29.6	$\pm$	0.5	&	11 &  0.846 \\
C6	&	2007.18	$\pm$	0.14	&	395.8	$\pm$	117.5	&	19.8 $\pm$	5.9	&	-255.2	$\pm$	0.6	&	7 &  0.945 \\
C7	&	2007.33	$\pm$	0.05	&	406.4	$\pm$	21.9	&	20.3 $\pm$	1.1	&	-24.7	$\pm$	4.0	&	11 &  0.961 \\
C8	&	2008.10	$\pm$	0.27	&	240.7	$\pm$	94.1	&	12.1 $\pm$	4.7	&	-210.0	$\pm$	4.9	&	8 &  0.882 \\
C9	&	2008.32	$\pm$	0.08	&	534.2	$\pm$	137.2	&	26.7 $\pm$	6.9	&	-167.4	$\pm$	1.7	&	4 &  0.999 \\
C10	&	2008.49	$\pm$	0.08	&	344.3	$\pm$	45.9	&	17.2  $\pm$	 2.3	&	-201.9	$\pm$	18.1	&	11 &  0.931 \\
C11	&	2009.29	$\pm$	0.10	&	545.9	$\pm$	170.7	&	27.3	$\pm$	8.5	&	-198.1	$\pm$	0.4	&	4 &  0.994 \\
C12	&	2009.45	$\pm$	0.23	&	184.6	$\pm$	70.0	&	9.2	$\pm$	3.5	&	-224.8	$\pm$	1.8	&	7 &  0.898 \\
C13	&	2009.89	$\pm$	0.39	&	187.9	$\pm$	84.2	&	9.4	$\pm$	4.2	&	-205.2	$\pm$	0.8	&	9 &  0.849 \\
C14	&	2010.37	$\pm$	0.14	&	170.5	$\pm$	21.8	&	8.5	$\pm$	1.1	&	-200.4	$\pm$	3.8	&	18	&  0.943 \\
C15	&	2011.41	$\pm$	0.16	&	247.3	$\pm$	39.3	&	12.4	$\pm$	2.0	&	-193.6	$\pm$	6.4	&	8 &  0.962 \\
C16	&	2011.46	$\pm$	0.25	&	115.7	$\pm$	20.1	&	5.8	$\pm$	1.0	&	-182.8	$\pm$	3.3	&	11	&  0.906 \\
C17	&	2012.16	$\pm$	0.45	&	106.9	$\pm$	50.0	&	5.4	$\pm$	2.5	&	-76.2	$\pm$	9.1	&	9	&  0.910 \\
C18	&	2013.08	$\pm$	0.26	&	160.5	$\pm$	61.5	&	8.0	$\pm$	3.1	&	-110.7	$\pm$	7.9	&	7	&  0.878 \\
C19	&	2013.68	$\pm$	0.57	&	190.5	$\pm$	164.7	&	9.5	$\pm$	8.2	&	-60.7	$\pm$	1.2	&	3	&  0.982 \\
C20	&	2014.28	$\pm$	0.09	&	409.3	$\pm$	129.7	&	20.5	$\pm$	6.5	&	-68.2	$\pm$	2.6	&	4	&  0.931 \\
C21	&	2014.45	$\pm$	0.34	&	223.8	$\pm$	113.2	&	11.2	$\pm$	5.7	&	-34.9	$\pm$	0.6	&	4	&  0.990 \\
C22	&	2015.32	$\pm$	0.05	&	686.0	$\pm$	132.0	&	34.3	$\pm$	6.6	&	-50.4	$\pm$	0.5	&	5	&  0.996 \\
C23	&	2015.48	$\pm$	0.05	&	642.3	$\pm$	114.8	&	32.2	$\pm$	5.7	&	-113.1	$\pm$	3.0	&	4	&  1.000 \\
C24	&	2015.69	$\pm$	0.10	&	305.9	$\pm$	26.3	&	15.3	$\pm$	1.3	&	-302.6	$\pm$	0.7	&	7 &  0.996 \\
C25	&	2015.84	$\pm$	0.11	&	263.5	$\pm$	58.1	&	13.2	$\pm$	2.9	&	-103.2	$\pm$	1.2	&	8 &  0.932 \\
C26	&	2016.00	$\pm$	0.10	&	293.1	$\pm$	40.5	&	14.7	$\pm$	2.0	&	-298.6	$\pm$	1.9	&	10	&  0.942 \\
C27	&	2016.57	$\pm$	0.05	&	446.4	$\pm$	26.0	&	22.3	$\pm$	1.3	&	-295.6	$\pm$	9.9	&	15	&  0.980 \\
C28	&	2016.69	$\pm$	0.07	&	306.1	$\pm$	23.4	&	15.3	$\pm$	1.2	&	-301.1	$\pm$	2.6	&	14 &  0.982 \\
C29	&	2017.04	$\pm$	0.09	&	270.6	$\pm$	17.8	&	13.5	$\pm$	0.9	&	-311.0	$\pm$	1.9	&	16 &  0.972 \\
C30	&	2017.51	$\pm$	0.22	&	272.2	$\pm$	98.0	&	13.6	$\pm$	4.9	&	-168.6	$\pm$	1.2	&	4	&  0.867 \\
C31	&	2017.71	$\pm$	0.13	&	262.6	$\pm$	27.8	&	13.1	$\pm$	1.4	&	-343.7	$\pm$	1.9	&	13	&  0.961 \\
C32	&	2018.05	$\pm$	0.24	&	241.1	$\pm$	131.9	&	12.1	$\pm$	6.6	&	-133.3	$\pm$	3.1	&	4	&  0.970 \\
C33	&	2018.69	$\pm$	0.17	&	285.4	$\pm$	163.1	&	14.3	$\pm$	8.2	&	-293.5	$\pm$	14.4	&	4	&  0.734 \\
C34	&	2019.41	$\pm$	0.13	&	135.1	$\pm$	7.3	&	6.8	$\pm$	0.4	&	-68.1	$\pm$	4.7	&	23	&  0.949 \\
C35	&	2020.11	$\pm$	0.18	&	127.3	$\pm$	12.4	&	6.4	$\pm$	0.6	&	-38.1	$\pm$	2.8	&	19	&  0.970 \\
C36	&	2020.47	$\pm$	0.69	&	66.0	$\pm$	21.8	&	3.3	$\pm$	1.1	&	-50.8	$\pm$	6.3	&	15	&  0.654 \\
        \hline 
	\end{tabular}
	\parbox{12.5cm}{\flushleft $^\mathrm{a}$ Number of epochs for which a given jet component was detected by our CE model fitting.\\}
\end{table*}

\section{Interferometric Data Set} \label{sec:ObsData}

In this work, we made use of interferometric images of AO\,0235+164 obtained at 15 GHz and available on the site that hosts the MOJAVE project\footnote{\url{https://www.cv.nrao.edu/MOJAVE/}} (Monitoring Of Jets in Active galactic nuclei with VLBA Experiments; e.g, \citealt{ 2009AJ....138.1874L}). In the case of observations at 43 GHz, radio maps were downloaded from the public database hosted by the Blazar group at Boston University\footnote{\url{https://www.bu.edu/blazars/VLBAproject.html}} {\bf \citep{2017ApJ...846...98J, 2022ApJS..260...12W}}. 

In total, we gathered 203 public FITS files containing 162 observations of AO\,0235+164 at 43 GHz obtained between July 2007 and April 2023, and 41 radio maps at 15 GHz from July 1995 to March 2009. All these images had their original field of view narrowed with the aim to reduce the computational costs involved in the Cross-Entropy global optimisation technique used in their model fittings \citep[e.g.,][]{2014MNRAS.441..187C}. 

From the best of our knowledge, the incorporation of both interferometric data into the analyses conducted in this work makes our investigation about the kinematic behaviour of the parsec-scale jet of AO\,0235+164 the most comprehensive from a temporal standpoint ever documented in the literature. 

\section{Results} \label{sec03}

We assumed that the individual radio maps of AO\,0235+164 can be decomposed into elliptical two-dimensional Gaussian components, where each of them depends on six independent parameters: peak intensity, $I_0$, two-dimensional peak position ($x_0, y_0$), with the coordinates $x$ and $y$ oriented respectively to right ascension and declination directions, semi-major axis, $a$, eccentricity, $\epsilon=\sqrt{1-(b/a)^2}$, where $b$ is the semi-minor axis, and the position angle of the major axis, $\psi$, measured positively from west to north. 

The structural parameters of the Gaussian components were determined using our Cross-Entropy (CE) global optimisation method \citep[e.g.,][]{RUBINSTEIN199789, 2009MNRAS.399.1415C}, already applied to other blazars in the context of interferometric radio images \citep{2014MNRAS.441..187C, 2017ApJ...851L..39C, 2021MNRAS.509.1646S, 2024ApJ...965....9N}. Following the criteria proposed by \citet{2014MNRAS.441..187C}, we determined the optimal number of Gaussian components in each one of the 203 images of AO\,0235+164 analysed in this work. Those procedures detected a single component (core) in a few epochs, not exceeding five simultaneous Gaussian components (core plus four jet components) during the approximately 28 years of monitoring covered in this work.

To assess the reliability of those elliptical Gaussian components in the $uv$-plane, we retrieved the corresponding calibrated complex visibility data of AO\,0235+164 from the public archives of the MOJAVE project and the Blazar Group at Boston University. These $uv$-data were modelled using the task MODELFIT in the DIFMAP software package (\citealt{1994BAAS...26..987S}) after feeding it with the CE-optimised values of the structural parameters of the Gaussian components. This allowed us to compute the reduced chi-squared values of the fits directly in the visibility domain, which are shown in \autoref{Chi-Square}. The median reduced chi-squared value across all 203 epochs analysed in this work is approximately 1.2\footnote{ \citet{2022ApJS..260...12W} fitted the complex visibility data of AO\,0235+164 between 2013 and 2018 using circular Gaussian components. Their model fittings resulted in a median reduced chi-square value of about 0.86, similar to the 0.96 found in our CE modellings in the $uv$-plane for the same interval.}, suggesting that our CE model-fitting in the image plane reasonably reproduces the complex visibility data of AO\,0235+164.

\subsection{Parsec-scale jet components in AO 0235+164} \label{sec:KinJetComp}

We have identified kinetically 36 non-static components in the parsec-scale jet of AO\,0235+164, all of them moving with constant proper motions from the core\footnote{We also fitted non-ballistic trajectories for those jet components with at least 9 epochs of detection, but no relevant differences between ballistic and non-ballistic fits were found considering the interval in which each of them is detected.}. Solid lines in the upper panel shown in \autoref{rxt_Fluxxt} represent linear regressions considering the equation

\begin{equation}
	r(t) = \mu\left(t-t_0\right) \; ,
	\label{eq:motionequation}
\end{equation}
\\where $r$ is the core-component distance at the time $t$, while $\mu$ and $t_0$ are respectively the apparent proper motion and the ejection epoch of the jet components.

The kinematic parameters of these 36 jet components are listed in \autoref{table:Kin_Param_AO0235_table}, together with the values of the Pearson correlation coefficient, $r_\mathrm{p}$, for the linear regressions shown in \autoref{rxt_Fluxxt}. The derived proper motions imply apparent speeds, $\beta_\mathrm{obs}$, that range from 3.0$c$ to 34.3$c$ at the source's redshift, indicating substantial variations in their individual proper motions. Similar apparent velocities were found by \citet{1996A&A...307...15C} through interferometric observations at 5 GHz from 1978.9 to 1983.5. \citet{2017ApJ...846...98J} estimated the apparent speeds for three jet components at 43 GHz, two of them compatible with our jet components C7 and C8. Using interferometric data at 43 GHz and based on the flare occurred between 2008 and 2010, \citet{2018MNRAS.475.4994K} identified a component with an apparent velocity  equivalent to $\beta_\mathrm{{app}} = (10.0\pm 1.6)c$, which is compatible with the component C8 identified in this work.

The instantaneous flux density of the jet components, $F_\nu$ is shown in the bottom panel of \autoref{rxt_Fluxxt}. This quantity is defined as

\begin{equation}
    F_\nu = 8 \ln{2} \left[ \frac{a^2 \sqrt{1 - \epsilon^2}}{\left(\Theta^\mathrm{FWHM}_{\mathrm{beam}}\right)^2 \sqrt{1 - \epsilon^2_{\mathrm{beam}}}} \right] I_0 \;.
    \label{eq:fluxdens}
\end{equation}
\\where $\Theta^{\mathrm{FWHM}}_{\mathrm{beam}}$ and $\epsilon_\mathrm{\,beam}$ are respectively the FWHM major axis and eccentricity of the synthesised elliptical \textsc{clean} beam of the observations.

The majority of the jet components exhibits either a systematic decrease of $F_\nu$ as time evolves (e.g., component C5), or an initial short phase where $F_\nu$ increases with time and a quasi-monotonic decrease of $F_\nu$ after reaching its maximum value (e.g., component C8). This behaviour is in agreement with predictions from the shock-in-jet models \citep[e.g.,][]{Konigl_1981, Marscher_Gear_1985, Hughes_et_al_1985, Valtaoja_et_al_1992, Turler_et_al_2000}.

\begin{figure}
	\includegraphics[width=0.95\columnwidth]{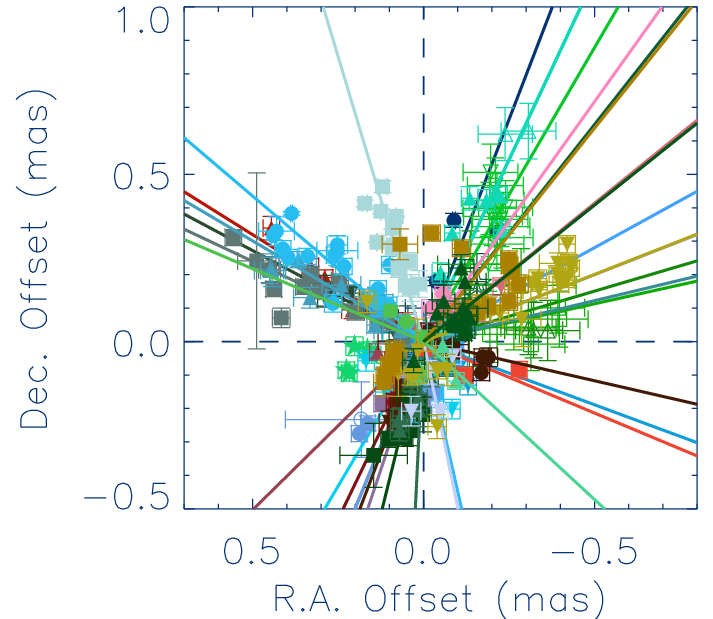}
        \centering
    \caption{Right ascension and declination offsets of the 36 jet components identified in this work. Coloured straight lines represent their respective mean position angles, while dashed horizontal and vertical grey lines are used to mark the four quadrants of the sky. Same colour scheme of \autoref{rxt_Fluxxt} is adopted here.}
   \label{plot_All_Rxt}
\end{figure}

The mean position angle of each jet component in AO\,0235+164, $\bar{\eta}$, also changes from component to component, reflecting on the spread over all quadrants on the plane of the sky shown in \autoref{plot_All_Rxt}. Notably, this unusual blazar jet behaviour had already been observed by \citet{2018MNRAS.475.4994K} in AO\,0235+164 using a shorter monitoring dataset (2007–2016). Similar findings were reported by \citet{2022ApJS..260...12W}, who analysed radio maps from 2007 to 2017, and by \citet{2024A&A...689A..56E} using data from 2008 to 2024.

\subsection{Limits for the Lorentz factor and the jet viewing angle of AO 0235+164} \label{subsec:MinLorentzFact_MaxJetViewAngle}

The minimum relativistic jet Lorentz factor, $\gamma_\mathrm{min}$, can be calculated from the maximum apparent speed among the jet components identified kinematically, $\beta_\mathrm{app}^\mathrm{max}$, through

\begin{equation}
    \gamma_{\mathrm{min}} = \sqrt{1 + \left(\beta_{\mathrm{app}}^{\mathrm{max}}\right)^2} \; ,
    \label{eq:betaappmin}
\end{equation}
\\ which implies $\gamma_{\mathrm{min}}= 34.3\pm6.6$ for the parsec-scale jet of AO\,0235+164 after using the apparent speed of the jet component C22 as a proxy for $\beta_\mathrm{app}^\mathrm{max}$.
This result is in good agreement with several previous estimates in the literature. \citet{Volvach2019} derived a value of approximately 35 by analysing time delays between radio and gamma-ray flares, whilst \citet{2017ApJ...846...98J} found values of $28.6\pm 6.9$ and $31.3 \pm 4.9$ for jet components B1 and B2, respectively. Similarly, \citet{2022ApJS..260...12W} reported Lorentz factors of $\gamma$ of $37.2 \pm 5.8$, $39.4 \pm 4.5$, and $31.6 \pm 4.0$ for components B1, B2, and B6. Earlier studies by \citet{1988ApJ...326..668O} and \citet{2000A&A...357...84Q} also proposed that $\gamma\ga 25$.

Nevertheless, lower values for $\gamma$ have also been suggested for AO\,0235+164. \citet{2015ARep...59..145V} reported a value of around 20, which is close to the $\gamma = \mathbf{16.8^{+3.6}_{-3.1}}$ found by \citet{10.1093/mnras/stad3250}. Furthermore, a value of $\gamma \approx 14$ was reported by both \citet{2018MNRAS.475.4994K} and \citet{2020ApJ...902...41W}.

On the other hand, we can also derive a very conservative maximum value for the jet viewing angle, $\phi_{\mathrm{max}}$, from the lowest apparent speed, $\beta_{\mathrm{app}}^{\mathrm{min}}$, among detected jet components \citep[e.g.,][]{2021MNRAS.509.1646S}

\begin{equation}
\phi_{\mathrm{max}} \approx \arccos{\left[\frac{\left(\beta_{\mathrm{app}}^{\mathrm{min}}\right)^{2} - 1}{\left(\beta_{\mathrm{app}}^{\mathrm{min}}\right)^{2} + 1}\right]} \; ,
\label{eq:theta_max}
\end{equation} 
\\leading to $\phi_{\mathrm{max}} =  36\fdg7\pm 7\fdg8$ 
(the apparent velocity of the jet component C3 was used in this calculation).

Estimates for the jet viewing angle of AO\,0235+164 vary in the literature, though most analyses suggest a value of less than $4\degr$. For instance, \citet{Volvach2015} derived a value of between $2\degr$ and $3\degr$, whilst a subsequent analysis by \citet{Volvach2019} suggested a smaller angle of approximately $1\fdg6$. Consistent with these findings, \citet{2011ApJ...735L..10A} established an upper limit of $2\fdg4$.

Other analyses have indicated even smaller angles. \citet{2009A&A...494..527H} found $\phi=0\fdg4$ (based on $\beta_\mathrm{app} = 2c$ from \citealp{2018MNRAS.475.4994K}), and \citet{2018MNRAS.475.4994K} themselves reported a value of $1\fdg7$. \citet{10.1093/mnras/stad3250} reported a value of $1\fdg42^{+1\fdg07}_{-0\fdg52}$. Furthermore, kinematic analysis of individual jet components has yielded a range of values from $0\fdg2\pm 0\fdg1$ to $3\fdg2\pm 0\fdg7$ \citep{2017ApJ...846...98J, 2022ApJS..260...12W}.

In contrast to these smaller angles, \citet{2020ApJ...902...41W} derived a much larger viewing angle of approximately $6\degr$ from their helical model for jet polarisation. However, the same study also derived $\phi=1\fdg7$ when assuming an apparent speed of $\beta_\mathrm{app} = 10c$.

\subsection{The brightness temperature of the core region} \label{sec:TBcore}

The brightness temperature of the core at the rest frame of the source, $T_\mathrm{B, rest}$, can be calculated from

\begin{equation}
	T_\mathrm{B, rest} = (1 + z) \, \frac{2 \, \ln{2}}{\pi k} \frac{c^2}{\nu^2} \frac{F_\nu}{a_{\mathrm{FWHM}} \, b_{\mathrm{FWHM}}} \; ,
	\label{eq:brightness_temperature_equation}
\end{equation}
\\where $k$ is the Boltzmann constant, and $a_{\mathrm{FWHM}}$ and $b_{\mathrm{FWHM}}$ are, respectively, the FWHM of the elliptical Gaussian components along the major and the minor axes. If the emitting region is unresolved angularly, the term $a_{\mathrm{FWHM}} \, b_{\mathrm{FWHM}}$ in \autoref{eq:brightness_temperature_equation} must be replaced by $d_{\mathrm{min}}^2$, define as \citep{Lobanov_2005}:

\begin{equation}
	d_{\mathrm{min}} = \frac{2^{2-\varpi/2}}{\pi} \left[ \pi \ln{2} \; \left(\Theta^\mathrm{FWHM}_{\mathrm{beam}}\right)^2 \sqrt{1 - \epsilon^2_{\mathrm{beam}}} \, \ln{\left( \frac{\mathrm{SNR}}{\mathrm{SNR} - 1} \right)} \right] ^{1/2} \;,
	\label{eq:dmin}
\end{equation}
\\where SNR is the signal-to-noise ratio and $\varpi$ is an index describing the weighting method used to generate the radio maps ($\varpi=0$ for uniform weighting and $\varpi=2$ for natural weighting; \citealt{Lobanov_2005}).

\begin{figure}	
    \includegraphics[width=1.0\columnwidth]{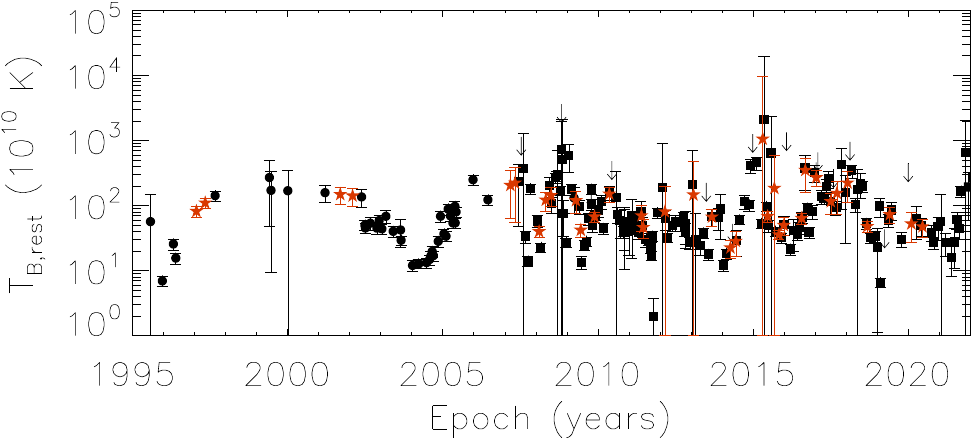}
        \centering
    \caption{Brightness temperature of the core at the rest frame of AO\,0235+164 as a function of time. Black circles and squares refer respectively to estimates at 15 and 43 GHz. Downward arrows correspond to upper limits (derived from \autoref{eq:dmin}), while red stars represent the interpolated values of $T_\mathrm{B,rest}$ at the ejection epochs of the jet components.}
   \label{fig:TBrest}
\end{figure}

The time behaviour of the brightness temperature of the core region derived from \autoref{eq:brightness_temperature_equation} is shown in \autoref{fig:TBrest}. We obtained for the core region at 15 GHz $0.07\leq T_\mathrm{B,rest}(10^{12}$ K) $\leq 2.85$, while for the data at 43 GHz, $0.07\leq T_\mathrm{B,rest}(10^{12}$ K) $\leq 21.88$. These values are compatible with those estimated in previous works \citep[e.g.,][]{2016ApJ...826..135L, 2021ApJ...923...67H}, confirming the fact that AO\,0235+164 can reach $T_\mathrm{B,rest}$ as higher as 10$^{13}$ K \citep[][]{2000PASJ...52..975F, 2018MNRAS.475.4994K, 2018ApJ...866..137L}.

\section{Discussion} \label{sec:Disc}

\subsection{Changes in the apparent speed and brightness temperature} \label{sec:TimeChangeVar}

We presented in \autoref{sec:KinJetComp} our results concerning the determination of the kinematic parameters of the parsec-scale jet components of AO\,0235+164. As mentioned previously, their apparent speeds and position angles exhibit substantial changes during all the monitoring time analysed in this work. A pertinent question is which mechanism could be responsible for those changes? To answer it, it is necessary to recover the relationship between $\beta_\mathrm{obs}$ and the jet viewing angle, $\phi$,

\begin{equation}
\beta_\mathrm{obs} = \frac{\beta \sin \phi}{(1-\beta \cos \phi)},
\label{eq:betaapp}
\end{equation}
\\where $\beta = v/c$, $v$ is the jet bulk speed. Introducing the relativistic Doppler boosting factor

\begin{equation}%
	\delta = \left[\gamma-\left(\gamma^2-1\right)^{1/2}\cos\phi\right]^{-1} \; ,
	\label{eq:doppler_factor}
\end{equation}
\\where $\gamma=(1-\beta^2)^{-1/2}$ is the jet Lorentz factor, we can rewrite \autoref{eq:betaapp} as

\begin{equation}
\beta_\mathrm{obs} = \delta\sqrt{\gamma^2-1}\sin \phi.
\label{eq:betaappnew}
\end{equation}

From \autoref{eq:betaappnew}, it is clear that any change in $\beta_\mathrm{obs}$ implies a time-variable $\gamma$ and/or $\phi$.

In addition, the rest-frame brightness temperature of the core region also presents variability, as shown in \autoref{fig:TBrest}. The relationship between $T_\mathrm{B,rest}$ and the intrinsic brightness temperature, $T_\mathrm{B,int}$, is stated as \citep[e.g.,][]{Readhead_1994, Kovalev_et_al_2005, Homan_et_al_2006}: 

\begin{equation}
T_\mathrm{B,rest} = \delta T_\mathrm{B,int},
\label{eq:TBrestTBint}
\end{equation}
\\where $T_\mathrm{B,int}$ is related to the equipartition temperature, $T_\mathrm{B,eq}(\simeq 5\times10^{10}$ K; \citealt{Readhead_1994}), through

\begin{equation}
T_\mathrm{B,int} = \zeta^{2/17} T_\mathrm{B,eq},
\label{eq:TBintTBeq}
\end{equation}
\\where $\zeta = u_\mathrm{p}/ u_\mathrm{B}$, the ratio between the energy density of the radiating particles, $u_\mathrm{p}$, and the energy density of the magnetic field, $u_\mathrm{B}$.

Substituting \autoref{eq:TBintTBeq} into \autoref{eq:TBrestTBint}, we obtain

\begin{equation}
T_\mathrm{B,rest} = \delta \zeta^{2/17} T_\mathrm{B,eq}.
\label{eq:TBrestTBeq}
\end{equation}

From \autoref{eq:TBrestTBeq}, we can realize that temporal changes in $\delta$ and/or $\zeta$ must drive variations in $T_\mathrm{B,rest}$. In the former possibility, changes in $\gamma$ and/or $\phi$ must occur, similar to what was inferred in the case of $\beta_\mathrm{obs}$.

Isolating $\delta$ in \autoref{eq:TBrestTBeq} and substituting it into \autoref{eq:betaappnew}, we obtain

\begin{equation}
 T_\mathrm{B,eq}\frac{\beta_\mathrm{obs}}{T_\mathrm{B,rest}} = \frac{\sqrt{\gamma^2-1}}{\zeta^{2/17}}\sin \phi.
\label{eq:TeqbetaobsdivTBrest}
\end{equation}

Note that the left-hand side of \autoref{eq:TeqbetaobsdivTBrest} put together quantities that can be estimated (almost) directly from the interferometric data of AO\,0235+164, while the right-hand side involves intrinsic quantities of the jet inlet region that are substantially harder to be inferred by the same data set. Note also that \autoref{eq:TeqbetaobsdivTBrest} provides an additional probe for signatures of a possible parsec-scale jet precession in AO\,0235+164, as we will discuss in the next section.

\subsection{A precessing jet in AO 0235+164}
\label{sec:PrecMod} 

Precession of the parsec-scale jet has already been invoked previously in the case of AO\,0235+164, not only to explain quasi-periodic variability in its continuum (e.g., \citealt{1999A&A...344..807K, 2015ARep...59..145V, 2019CosRe..57...85V, 2023MNRAS.518.5788O}), but also to deal with changes in the jet position angle during decades of interferometric monitoring of this source at radio wavelengths (e.g., \citealt{1997AcASn..38..397Z, 1999Ap&SS.266..495C, 2024A&A...689A..56E}). Here, we revisited the scenario of the parsec-scale jet precession of AO\,0235+164, using observational constraints from the longest interferometric monitoring of this source, as well as changes in its brightness temperature during this period.

The kinematic precession model adopted in this work was previously applied to other blazars \citep[e.g.][]{1999A&A...344...61A, 2000A&A...355..915A, 2004MNRAS.349.1218C, 2004ApJ...602..625C, 2013MNRAS.428..280C, 2017ApJ...851L..39C}. It considers a relativistic jet with constant speed $v=\beta c$ ($0\leq \beta <1$), forming an angle $\phi$ in relation to the line of sight which is timely variable due to the precession motion of the jet. The jet precession occurs with a period $P_\mathrm{prec,obs}$ inferred in the observer's reference frame, forming an apparent cone shape in which its axis forms an angle $\phi_0$ with respect to the line of sight. This apparent precession cone has a semi-opening angle $\varphi_{0}$ (an schematically view is shown in figure 1 in \citealt{2009MNRAS.399.1415C}). Thus, precession induces periodic variations in $\phi$ that is quantified as (e.g., \citealt{2017ApJ...851L..39C})

\begin{equation} \label{cosphi}
\cos\phi(\tau_\mathrm{s})=-e_\mathrm{x,s}(\tau_\mathrm{s})\sin\phi_0+\cos\varphi_0\cos\phi_0,
\end{equation}
\\where $\tau_\mathrm{s} = t_\mathrm{s}/P_\mathrm{prec,s}$, $t_\mathrm{s}$ is the inferred time at the source's reference frame, and $e_\mathrm{x,s}(\tau_\mathrm{s})$ is given as

\begin{equation}\label{exs}
e_\mathrm{x,s} (\tau_\mathrm{s}) = \sin \varphi_\mathrm{0} \sin (\iota2\pi \Delta \tau_\mathrm{s}),
\end{equation}
\\where $\iota$ is the precession direction ($\iota = 1$ for clockwise and $\iota = -1$ for counterclockwise sense), $\Delta \tau_\mathrm{s} = \tau_\mathrm{s} - \tau_\mathrm{0,s }$, and $\tau_\mathrm{0,s} = t_\mathrm{0,s}/P_\mathrm{prec,s}$ is the precession phase (see \citealt{2009MNRAS.399.1415C} for more details).

A varying $\phi$ due to jet precession also introduces periodic modulations to the apparent velocities of the jet components (\autoref{eq:betaapp}), the relativistic Doppler boosting factor associated with the underlying jet (\autoref{eq:doppler_factor}) and the quantity $T_\mathrm{B,eq}\frac{\beta_\mathrm{obs}}{T_\mathrm{B,rest}}$ (\autoref{eq:TeqbetaobsdivTBrest}), as well as to the jet's instantaneous position angle on the plane of the sky, $\eta$ (e.g., \citealt{2017ApJ...851L..39C}):

\begin{equation}
\tan \eta(\tau_\mathrm{s}) = \frac{A(\tau_\mathrm{s}) \sin \eta_\mathrm{0} + e_\mathrm{y,s}(\tau_\mathrm{s}) \cos \eta_\mathrm{0}}{A(\tau_\mathrm{s}) \cos \eta_\mathrm{0} - e_\mathrm{y,s}(\tau_\mathrm{s}) \sin \eta_\mathrm{0}},
\label{eq:04}
\end{equation}
\\where $A(\tau_\mathrm{s})$ and $e_\mathrm{y,s}(\tau_\mathrm{s})$ are defined respectively as

\begin{equation} \label{A_tau}
A(\tau_\mathrm{s})=e_{x,\mathrm{s}}(\tau_\mathrm{s})\cos\phi_0+\cos\varphi_0\sin\phi_0,
\end{equation}
\\and

\begin{equation}\label{eys}
e_\mathrm{y,s}(\tau_\mathrm{s}) = \sin \varphi_\mathrm{0} \cos (\iota2\pi \Delta \tau_\mathrm{s}).
\end{equation}

The inferred time interval in the observer's reference frame, $\Delta t_\mathrm{obs}$, is related to the elapsed time in the source's reference frame as

\begin{equation} \label{elapsed_time}
\frac{\Delta t_\mathrm{obs}}{P_\mathrm{prec,obs}}=\frac{\int_{0}^{\Delta\tau_\mathrm{s}}\delta^{-1}(\tau)d\tau}{\int_0^1\delta^{-1}(\tau)d\tau}.
\end{equation}

In our purely kinematic precession model there are seven free parameters to be determined: $\iota$, $P_\mathrm{prec,obs}$, $\gamma$, $\eta_0$, $\varphi_0$, $\phi_0$ and $\tau_\mathrm{0,s}$. The last six free parameters were obtained via the CE method \citep[][]{2009MNRAS.399.1415C, 2013MNRAS.428..280C} for both values of $\iota$, and after fixing the range for the possible values that each of the free model parameters can assume. Regarding the jet bulk Lorentz factor, the CE optimization assumed values between 27.75 ($\beta=0.999351$, corresponding to $\gamma_\mathrm{min}$ minus its one-sigma uncertainty) and 50.00 ($\beta=0.9998$). For $P_\mathrm{prec,obs}$, values ranged from 2.0 to 30 years, while the range between 0\degr to 40\fdg5 ($\phi_\mathrm{max}$ plus its one-sigma uncertainty) was considered for $\phi_0$ and $\varphi_0$\footnote{During the whole CE optimization, if a given tentative solution has $\phi_0 + \varphi_0 > \phi_\mathrm{max}$, both angles are automatically replaced to preserve the condition $\phi\leq\phi_\mathrm{max}$ mentioned in \autoref{subsec:MinLorentzFact_MaxJetViewAngle}.}. 
 
For the remaining free parameters, we assumed $-360\degr < \eta_0\leq 0\degr$, since the jet components of AO\,0235+164 spread across all sky quadrants and $0\leq\tau_\mathrm{0,s}<1$.

Having fixed the precession direction ($\iota$), our CE method minimizes a merit function $S(k)$ at iteration $k$, which is defined as

\begin{equation}
S(k)=S_1(k)+S_2(k)+S_3(k),
\label{eq:10}
\end{equation}
\\where the individual terms $S_1(k)$, $S_2(k)$ and $S_3(k)$ are written as

\begin{equation} \label{merit_function1}
S_1(k) = -\ln\left\{\prod\limits_{i=1}^{N_\mathrm{d}}  \frac{\exp\left[-\frac{1}{2}\left(S_{\alpha_i}^2(k)+S_{\delta_i}^2(k)+S_{r_i}^2(k)\right)\right]}{\left(2\pi\right)^{3/2}\sigma_{\Delta\alpha_i}\sigma_{\Delta\delta_i}\sigma_{\Delta r_i}}\right\},
\end{equation}

\begin{equation} \label{merit_function2}
S_2(k) = -\ln\left\{\prod\limits_{i=1}^{N_\mathrm{kin}}  \frac{\exp\left[-\frac{1}{2}\left(S_{\beta_{\mathrm{obs},i}}^2(k)+S_{\eta_i}^2(k)+S_{\beta_{\mathrm{obs,}i}^\prime}^2(k)\right)\right]}{\left(2\pi\right)^{3/2}\sigma_{\beta_{\mathrm{obs},i}}\sigma_{\eta_i}\sigma_{\beta_{\mathrm{obs,}i}^\prime}}\right\}, 
\end{equation}
\\and

\begin{equation} \label{merit_function3}
S_3(k) = -\ln\left\{\prod\limits_{i=1}^{N_\mathrm{TB}}  \frac{\exp\left[-\frac{1}{2}S_\mathrm{TB_i}^2(k)\right]}{(2\pi)^{1/2}\sigma_\mathrm{TB_i}}\right\}.
\end{equation}

In the case of \autoref{merit_function1}, $N_\mathrm{d}$ is the total number of the Gaussian jet components found during the whole monitoring interval analysed in this work, and $\sigma_{\Delta \alpha_i}$, $\sigma_\mathrm{\Delta \delta_i}$, and $\sigma_\mathrm{\Delta r_i}$ correspond respectively the uncertainties in right ascension, declination, and core-component distance. Besides, $S_{\alpha_i}(k)= \sigma_{\Delta\alpha_i}^{-1}\left[\Delta\alpha_i-\Delta\alpha_\mathrm{mod_i}(k)\right]$, $S_{\delta_i}(k)= \sigma_{\Delta\delta_i}^{-1}\left[\Delta\delta_i-\Delta\delta_\mathrm{mod_i}(k)\right]$,  $S_{r_i}(k)= \sigma_{\Delta r_i}^{-1}\left[\Delta r_i-\Delta r_{\mathrm{mod}_i}(k)\right]$, $\Delta r_i^2 = \Delta\alpha_i^2+\Delta\delta_i^2$, $\Delta r_{\mathrm{mod}_i}^2 = \Delta\alpha_{\mathrm{mod}_i}^2+\Delta\delta_{\mathrm{mod}_i}^2$, $\Delta\alpha_i$ and $\Delta\delta_i$ correspond to the displacements of the identified components of the jet $i$ in relation to the core component, and $\Delta \alpha_\mathrm{mod_{i}}$ and $\Delta \delta_\mathrm{mod_{i}}$ correspond to the displacements of the identified jet components $i$ predicted by the precession model (see equations 12 and 13 in \citealt{2013MNRAS.428..280C} for the exact definitions of $\Delta \alpha_\mathrm{mod_{i}}$ and $\Delta \delta_\mathrm{mod_{i}}$).

In \autoref{merit_function2}, $N_\mathrm{kin}$ is the number of jet components kinetically identified in this work, $S_{\beta_{\mathrm{obs},i}}(k)= \sigma_{\beta_{\mathrm{obs},i}}^{-1}\left[\beta_{\mathrm{obs},i}-\beta_{\mathrm{obs},i}^\mathrm{mod}(k)\right]$, $S_{\eta_i}(k)= \sigma_{\eta_i}^{-1}\tan\left[\eta_{\mathrm{obs},i}-\eta_{\mathrm{obs},i}^\mathrm{mod}(k)\right]$, $S_{\beta_{\mathrm{obs,}i}^{\prime}}(k)= \sigma_{\beta_{\mathrm{obs,}i}^\prime}^{-1}\left[\beta_{\mathrm{obs},i}^{\prime}-\beta_{\mathrm{obs},i}^\mathrm{mod\prime}(k)\right]$, $\beta_{\mathrm{obs},i}$ and $\eta_{\mathrm{obs},i}$ correspond respectively to the apparent velocity and position angle of the jet components on the plane of sky, to which $\sigma_{\beta_{\mathrm{obs},i}}$ and $\sigma_{\eta_i}$ correspond to their respective uncertainties. Primed and non-primed quantities in \autoref{merit_function2} are given in terms of time (in the observer's reference frame) and position angle, respectively, with $\sigma_{\beta_{\mathrm{obs,}i}^\prime} = \sqrt{\sigma_{\beta_{\mathrm{obs},i}}\sigma_{\eta_i}}$.

\begin{table*} 		
\caption{Precession model parameters optimized by our CE technique for clockwise and counter-clockwise senses of precession.}\label{tab:PrecParams}	
\begin{tabular}{@{}ccccccccccc@{}}
\hline
$\iota$ & $P_\mathrm{prec,obs}$ & $\gamma$ & $\eta_0$ & $\phi_0$ & $\varphi_0$ & $\tau_\mathrm{0,s}$ & $\zeta$ & $S(k_\mathrm{max})$\\
 & (years) &  & (deg) & (deg) & (deg) &  & ($10^5$) & ($10^4$)\\
\hline
1 & 8.39 $\pm$ 0.24  &  29.2 $\pm$ 3.9  &     -26.7 $\pm$ 8.0   &   7.9 $\pm$ 2.6   &       9.3 $\pm$ 2.5   &   0.03 $\pm$ 0.51  & $1.0\pm0.5$  &  3.30 \\
-1 & 6.00 $\pm$ 0.09   &   29.8 $\pm$ 4.0   &    -348.7 $\pm$ 12.9   &   6.2 $\pm$ 1.5   &       9.0 $\pm$ 0.8   &     0.79 $\pm$ 0.29 & $1.0\pm 0.3$  & 6.42 \\
\hline
\end{tabular}

$\textbf{Notes}$: The uncertainties in each parameter are at 1$\sigma$-level; Merit function in the final iteration ($k_\mathrm{max} = 65$) is shown in the ninth column of this table.
\end{table*}

\begin{figure*}
	\includegraphics[width=0.88\textwidth]{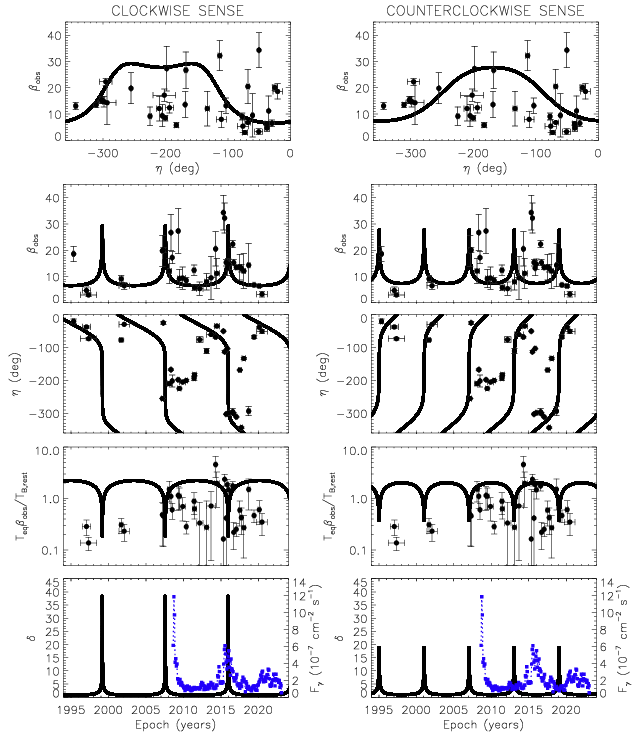}
        \centering
    \caption{Jet precession models for AO\,0235+164 obtained via the CE method considering clockwise and counter-clockwise senses of precession (left and right panels, respectively). Solid lines represent the predictions of the precession models given in Table 7 in the time-independent $\beta_\mathrm{obs}-\eta$ plane (top panels), while time behaviour of these two quantities are shown in the middle panels. Black circles refer to the kinematic parameters of the jet components listed in \autoref{table:Kin_Param_AO0235_table}.
   Bottom panels show the temporal behaviour of the dimensionless quantity $T_\mathrm{B,eq}\beta_\mathrm{obs}T_\mathrm{B,rest}^{-1}$, and the Doppler factor (black line) superimposed on the public {\it Fermi} $\gamma$-ray light curve (blue squares).}
   \label{fig:Prec}
\end{figure*}

Motivated by \citet[]{2021MNRAS.509.1646S}, who analysed the instantaneous behaviour of the core brightness temperature in other two blazars, we have included an additional observational constraint in relation to our previous works on jet precession. This third constraint is encoded in the term $S_3$ defined in \autoref{merit_function3}. 
The term $S_3$ involves $T_\mathrm{eq}$, $\beta_\mathrm{obs}$, and $T_\mathrm{B,rest}$, estimated at the moments of jet component ejections (red points in \autoref{fig:TBrest} calculated from spline interpolations). These three quantities are in the left side of \autoref{eq:TeqbetaobsdivTBrest}, remaining on its right side the parameters $\gamma$, $\zeta$ and $\phi$. If these three last parameters do not change in time, $T_\mathrm{B,eq}\beta_\mathrm{obs}T_\mathrm{B,rest}^{-1}$ is constant too, which is false in the case of AO\,0235+164, as shown in \autoref{fig:Prec}. Our kinematic jet precession model assumes a constant $\gamma$, such that changes in $T_\mathrm{B,eq}\beta_\mathrm{obs}T_\mathrm{B,rest}^{-1}$ must be attributed to $\zeta$ and/or $\phi$. We have assumed that $\zeta$ remains constant throughout the period considered in this work to simplify our precession modellings. This assumption made us consider $\zeta$ as the eighth free parameter in our precession model, to be optimised by the CE technique. During CE optimisations, $\zeta$ was allowed to vary from $10^{-10}$ ($u_\mathrm{p} << u_\mathrm{B}$) to $10^5$ ($u_\mathrm{p} >> u_\mathrm{B}$). 

The results from the CE optimisation for both clockwise and counter-clockwise senses of jet precession are shown in \autoref{fig:Prec}, while the respective optimized precession parameters are given in \autoref{tab:PrecParams}. The parameters $\gamma$, $\phi_0$ and $\varphi_0$ obtained at clockwise and counter-clockwise senses agree each other considering their respective uncertainties. The precession period at the observer's reference frame for the clockwise sense corresponds to 8.39 years, a slightly longer than 6.00 years found in the case of counter-clockwise precession. The precession period of 8.4 years is compatible with the periodicity of about 8 years detected in the optical light curve of AO\,0235+164 \citep{2006A&A...459..731R, 2016Galax...4...17F, 2017ApJ...837...45F, 2022MNRAS.513.5238R}. On the other hand, the counter-clockwise precession period of 6.0 years is compatible with a periodicity of 5-6 years identified at radio frequencies \citep{2000ApJ...545..758R, 2006ApJ...650..749L, 2007ASPC..373..195F, 2021MNRAS.501.5997T} and optical bands \citep{2001A&A...377..396R, 2002A&A...381....1F}. 

Both jet precession models reasonably fit the observational data set, with the clockwise precession being slightly favoured when considering the final value of the merit function, $S(k_\mathrm{max})$, as shown in \autoref{tab:PrecParams}. Coincidence between the two main gamma-ray flares seen in the public {\it Fermi} light curve of AO\,0235+164 and the peaks of the Doppler factor predicted by the clockwise jet precession represents an additional advantage over the counter-clockwise precession (see the bottom panels in \autoref{fig:Prec}). However, the values of $S(k_\mathrm{max})$ differ by less than a factor of two, which implies the counter-clockwise sense of precession for AO\,0235+164 cannot be ruled out at all.

Although jet precession can roughly provide the correct amplitude variation of $T_\mathrm{B,eq}\beta_\mathrm{obs}T_\mathrm{B,rest}^{-1}$, additional fluctuations are seen in \autoref{fig:Prec}, suggesting that $\zeta$ might have also varied during this interval. Indeed, it is perfectly acceptable since a switch between low and high brightness temperature states must be related to changes in the energy densities of the particles and/or the magnetic field (e.g., \citealt{Homan_et_al_2006}).

Some jet components with apparent velocities of about $10c$ and position angles around -200\degr are systematically below the curves representing our precession models in the $\beta_\mathrm{obs}-\eta$ plots shown in \autoref{fig:Prec}. Additional oscillations superimposed on the jet precession might explain this discrepancy (see \autoref{subsec:NoddingMotion}).

\section{A supermassive binary black hole system in AO 0235+164}
\label{sec04} 

The existence of a SMBHB in the nucleus of AO\,0235+164 has been proposed in the literature, based mainly on the quasi periodic variability detected in its radio and optical continuum light curves (e.g., \citealt{1999Ap&SS.266..495C, 2006ApJ...650..749L, 2015ARep...59..145V, 2015ARep...59..851B, 2017Ap&SS.362...99W, 2019CosRe..57...85V, 2022MNRAS.513.5238R, 2023MNRAS.518.5788O}). 

\citet{1999Ap&SS.266..495C} estimated a lower limit of $1.46\times 10^8$ M$_\odot$ for the total mass of the SMBHB in AO\,0235+164 after interpreting the periodicities of 1.81 and 3.63 years found in its single-dish flux density at 5 GHz. \citet{2006ApJ...650..749L} found six quasi-periodic oscillations in the radio light curves of AO\,0235+164, attributing them to oscillatory accretion rates due to acoustic $p$-mode oscillations (e.g., \citealt{2003MNRAS.344..978R, 2005MNRAS.357L..31R}) in a geometrically thick disc probably induced by a SMBHB with a total mass of $4.7\times 10^8$ M$_\odot$. \citet{2015ARep...59..145V} and \citet{2019CosRe..57...85V} found periodicities of about two and eight years in a multi-wavelength analysis of AO\,0235+164, which were interpreted respectively as the orbital and jet precession periods in a SMBHB of $\sim10^{10}$ M$_\odot$. \citealt{2022MNRAS.513.5238R} suggested a binary system of supermassive black holes as an explanation for the major flares in the $R$-band light curve of AO\,0235+164 are double-peaked, with the secondary peak following the primary by about 2 years, a similar behaviour observed in the case of the blazar OJ\,287 (e.g., \citealt{1996ApJ...460..207L}).

In this section, we analyse the possibility of the parsec-scale jet precession in AO\,0235+164 is driven by a secondary supermassive black hole (SMBH) with mass $M_\mathrm{s}$ in a non-coplanar circular orbit of radius $d_\mathrm{BH}$ around a primary SMBH with mass $M_\mathrm{p}$ \citep[e.g.][]{1988ApJ...325..628S, 1997ApJ...478..527K, 2000A&A...355..915A, 2000A&A...360...57R, 2004ApJ...602..625C, 2004MNRAS.349.1218C, 2006ApJ...653..112C, 2013MNRAS.428..280C, 2013A&A...557A..85R, 2017ApJ...851L..39C, Nandi2021, 2024MNRAS.530.4902S}, assuming that its jet comes from the primary accretion disc torqued by the secondary SMBH. Some consequences of this scenario on the periodic variabilities found in the continuum emission of this source is also explored in this particular section. Note also that all the results presented hereafter are based only on the clockwise jet precession model discussed in \autoref{sec:PrecMod} without loss of generality, since the choice of the counter-clockwise jet precession parameters does not alter substantially the physical characteristics of the putative SMBHB in AO\,0235+164.

\begin{figure}
 	\includegraphics[width=1.0\columnwidth]{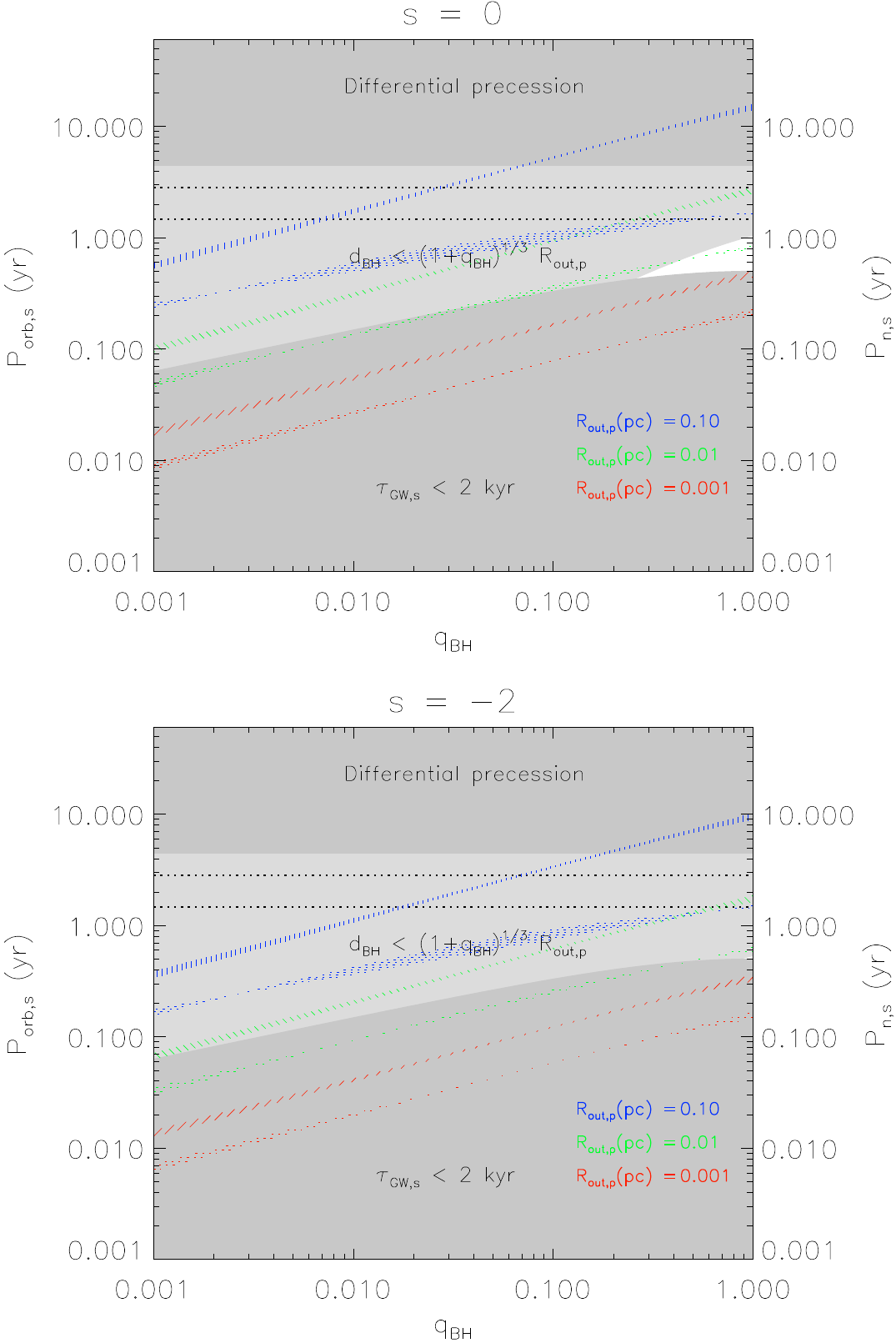}
        \centering
    \caption{Orbital period of the secondary BH and nodding period as a function of $q_\mathrm{BH}$ for a primary accretion disc with a constant surface density ($s=0$, top panel) and a decreasing surface density ($s=-2$, bottom panel) in the ballistic jet precession scenario ($P_\mathrm{prec,s}= 4.32$ yr). Slanted hatched stripes refer to the orbital period of the secondary BH considering outer radii for the primary disc of 0.1 pc (blue colour), 0.01 pc (green colour) and 0.001 pc (red colour) at $1\sigma-$level calculated from \autoref{PprecPorbSMBHB}. Dotted stripes show the behaviour of the nodding period for the same three values of the outer radii. The dotted horizontal lines correspond to the periodicities of 2.8 and 5.4 years detected in the radio and optical light curves of AO\,0235+164 converted to the source's reference frame. Dark gray regions delimitate the non-solid body precession regime and BH separations that lead to gravitational-wave time-scales in the source’s reference frame shorter than $2000$ years. The light gray region implies a BH separation that is smaller than about the outer radius of the primary disc. Viable solutions are found in the white region of the plots.}
   \label{fig:SMBHB_Periods_ballistic}
\end{figure}

\begin{figure}
 	\includegraphics[width=1.0\columnwidth]{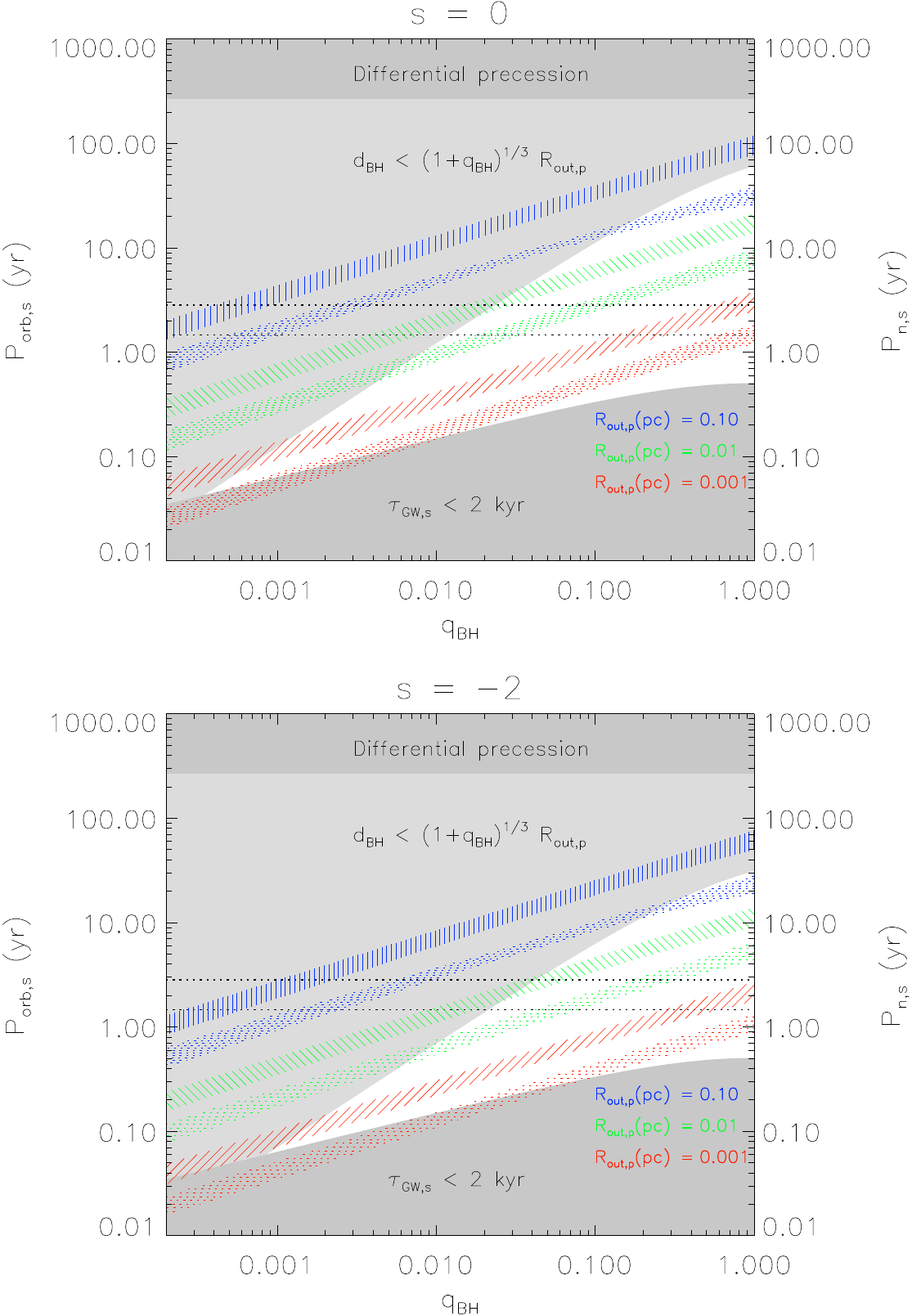}
        \centering
    \caption{ The same plots shown in \autoref{fig:SMBHB_Periods_ballistic} but considering a non-ballistic jet precession scenario ($P_\mathrm{prec,s}= 188$ yr).}
   \label{fig:SMBHB_Periods}
\end{figure}

\subsection{Physical parameters of the SMBHB}\label{subsec:PhysParam} 

Considering the formalism provided in \citet{2000MNRAS.317..773B}, as well as assuming an accretion disc with a power-law surface density distribution, we can write a relationship between the precession period of the accretion disc/jet, $P_\mathrm{prec,s}$, and the orbital period of the secondary BH, $P_\mathrm{orb,s}$, both given at the source's reference frame, as follows (e.g., \citealt{2024MNRAS.530.4902S})

\begin{equation}\label{PprecSMBHB}
\left[K(s)\cos\varphi_0\right]\left(\frac{P_\mathrm{prec,s}}{P_\mathrm{orb,s}}\right)=\left[\frac{\left(1+q_\mathrm{BH}\right)^{1/2}}{q_\mathrm{BH}}\right]\left(\frac{R_\mathrm{out,p}}{d_\mathrm{BH}}\right)^{-3/2},
\end{equation}
\\where $q_\mathrm{BH}=M_\mathrm{s}/M_\mathrm{p}$, $R_\mathrm{out,p}$ is the outer radius of the primary disc, and $K(s)\approx 0.19-0.47$ for a power-law surface density accretion disc with an index $s$ between 0 and -2 \citep{Larwood1996, 2000MNRAS.317..773B}.

Using the third Kepler's law, \autoref{PprecSMBHB} can be rearranged to obtain $P_\mathrm{orb,s}$ in terms of $q_\mathrm{BH}$ after fixing the values of $s$, $\varphi_0$, $P_\mathrm{prec,s}$ and $R_\mathrm{out,p}$:

\begin{equation}\label{PprecPorbSMBHB}
P_\mathrm{orb,s}=\sqrt{2\pi K(s)\cos\varphi_0 P_\mathrm{prec,s} \sqrt{\frac{R_\mathrm{out,p}^{3}}{GM_\mathrm{BH}}}\frac{q_\mathrm{BH}}{\sqrt{1+q_\mathrm{BH}}}},
\end{equation}
\\where $G$ is the gravitational constant, $M_\mathrm{BH}(=M_\mathrm{s}+M_\mathrm{p})$ is the total mass of the SMBHB.

We show in \autoref{fig:SMBHB_Periods_ballistic} the behaviour of $P_\mathrm{orb,s}$ as a function of $q_\mathrm{BH}$ for three generic values of $R_\mathrm{out,p}$: 0.1, 0.01 and 0.001 pc. In those calculations, it was assumed $M_\mathrm{BH}=10^{8.54\pm 0.12}$ M$_{\sun}$, the mean value considering the estimates found in the literature \citep{2000ChA&A..24....1F, 2006ApJ...650..749L, 2021ApJS..253...46P, 2023ApJS..265...14R}, and  $P_\mathrm{prec,s}= 4.32\pm 0.12$ yr, suitable for a ballistic jet precession scenario (e.g., \citealt{2004ApJ...615L...5R,1999A&A...344...61A, 2000A&A...355..915A, 2004MNRAS.349.1218C, 2004ApJ...602..625C}).

The precession period of the primary disc and its relationship with the orbital period of the misaligned secondary BH, as given in \autoref{PprecPorbSMBHB}, assumes that the disc precesses rigidly, implying that the orbital period of the secondary BH must be shorter than the precession period, as well as the former quantity must be longer than the orbital period at the outer radius of the primary disc, $P_\mathrm{out,p}$ \citep{2000MNRAS.317..773B}. The first requirement ($P_\mathrm{prec,s}>>P_\mathrm{orb,s}$) requires that acceptable orbital periods must be below of the upper rectangular dark grey region labelled as `Differential precession' in the two panels of the \autoref{fig:SMBHB_Periods_ballistic}. The second requirement ($P_\mathrm{orb,s}>>P_\mathrm{out,p}$) produces an exclusion region shown in the same panels as a light grey region labelled as `$d_\mathrm{BH} < \left(1+q_\mathrm{BH}\right)^{1/3}R_\mathrm{out,p}$'.

A third requirement that must be fulfilled is the stability of the binary system against gravitational wave losses. From the time-scale of the gravitational waves in the source's reference frame, $\tau_\mathrm{GW,s}$ (e.g., \citealt{1983bhwd.book.....S}), the shortest orbital period of the secondary SMBH, $P_\mathrm{orb,s}^\mathrm{min}$, to avoid the coalescence of the SMBHB within an interval of $\tau_\mathrm{GW,s}$ can be determined from

\begin{equation}\label{PprecPorbminGW}
P_\mathrm{orb,s}^\mathrm{min}\ge 2\pi\left[\frac{256}{5c^5}\left(GM_\mathrm{BH}\right)^{5/3}\frac{q_\mathrm{BH}}{\left(1+q_\mathrm{BH}\right)^2}\tau_\mathrm{GW,s}\right]^{3/8}.
\end{equation}

In \autoref{fig:SMBHB_Periods_ballistic}, the region where $\tau_\mathrm{GW,s}\le 2000$ yr is represented by the bottom dark grey zone labelled as `$\tau_\mathrm{GW,s}<$ 2 kyr'. This particular choice for $\tau_\mathrm{GW,s}$ implies in a relative change of $P_\mathrm{orb,s}$ of smaller than {\bf 2 percent} during the last 100 years, smaller than the estimated uncertainty for $P_\mathrm{prec,obs}$. It assures a time-steady accretion disc/jet precession in AO\,0235+164 independent of the chosen values of $s$ and $q_\mathrm{BH}$ during almost thirty years of interferometric monitoring of this source considered in this work.

The allowed combinations among $R_\mathrm{out,p}$, $s$ and $q_\mathrm{BH}$ that lead to $P_\mathrm{orb,s}$ compatible with the inferred jet precession period in AO\,0235+164 are found in the white region in \autoref{fig:SMBHB_Periods_ballistic}. We can realize that no reasonable solution has been found for $s=-2$\footnote{ Either $s\ga-1.8$ or $\tau_\mathrm{GW,s}\la 1.7$ kyr are necessary conditions to provide acceptable solutions considering the ballistic jet precession scenario.}. For $s=0$, orbital periods roughly between 0.5 and 1 yr are allowed in this source, implying in a separation between the primary and secondary black holes ranging from about 2 to 4 mpc (0.25-0.51 $\mu$as). Moreover, $R_\mathrm{out,p}$ must be ranged between 1 and 3 mpc, as well as $q_\mathrm{BH}\ga 0.3$.

The inferred range for the separation between the primary and secondary black holes in AO\,0235+164 are not angularly resolved by VLBI experiments currently available. However, a SMBHB is expected to emit gravitational waves at a frequency corresponding to twice the orbital frequency in a circular orbit (e.g., \citealt{1963PhRv..131..435P, 1972ARA&A..10..335P}). For $P_\mathrm{orb,s}$ between 0.5 and 1 yr estimated, gravitational waves are respectively generated between 31.7 and 15.8 nHz. Such frequencies are outside of the range planned to be covered by the Laser Interferometer Space Antenna (LISA; \citealt{2017arXiv170200786A}), but accessible for experiments like the Pulsar Timing Arrays (PTAs; e.g., \citealt{2023A&A...678A..48E} and references therein) in full operation. The detection of gravitational radiation in such frequency range could confirm the presence of a SMBHB in the core of AO\,0235+164.

The 36 jet components identified kinematically in this work exhibit proper motions that are consistent with ballistic displacements. However, several of these components (namely C5, C14, C16, C18, C27, C29, C34, C35 and C36) display a relatively high scatter around the receding motions predicted by the ballistic scenario. Although fitting non-ballistic trajectories to these components did not yield significant differences compared with the ballistic model over the interval during which they were observed by the VLBI experiments, their backward or forward extrapolations could result in measurable discrepancies between these two scenarios. It should be noted that accelerated motions may arise either from intrinsic variations in the jet Lorentz factor or from changes in the three-dimensional orientation of the jet as it propagates (e.g. \citealt{2009ApJ...706.1253H}).

In the case of jet precession involving non-ballistic motions, light-travel time effects shorten the precession period measured in the observer's reference frame (e.g., \citealt{2004ApJ...615L...5R}). Assuming that the inferred jet precession parameters in this work still hold, light-travel time effects implies $P_\mathrm{prec,s}= 188\pm 77$ yr, which was derived from (e.g., \citealt{2003ChJAA...3..513R, 2013MNRAS.428..280C, 2017ApJ...851L..39C})

\begin{equation} \label{Pprecobssource}
P_\mathrm{prec,s} = \frac{P_\mathrm{prec,obs}}{\left(1+z\right)\left(1-\beta\cos\varphi_0\cos\phi_0\right)},
\end{equation}
\\also considering the values of $P_\mathrm{prec,obs}$, $\varphi_0$ and $\phi_0$ listed in \autoref{tab:PrecParams} for the clockwise sense of precession.

Following the same approach described below, we present in \autoref{fig:SMBHB_Periods} the results considering the non-ballistic jet precession scenario for AO\,0235+164. As expected, the ranges for viable values of $P_\mathrm{orb,s}$ and $q_\mathrm{BH}$ clearly widen in comparison with those found assuming ballistic precession as shown in \autoref{fig:SMBHB_Periods_ballistic}, implying 0.05 yr$\la P_\mathrm{orb,s}\la$ 60 yr and $q_\mathrm{BH}\ga 0.0004$. Besides, $d_\mathrm{BH}$ must be between 0.0004 and 0.05 pc (0.05-6.32 $\mu$as), $R_\mathrm{out,p}$ between 0.0004 and 0.06 pc, and gravitational-wave frequencies between about 0.3 and 317 nHz.

\subsection{Nodding motions}\label{subsec:NoddingMotion} 

Besides the precession of the primary accretion disc in a SMBHB with the secondary BH revolving in a misaligned orbit,  \citet{1982ApJ...260..780K} showed that short-term oscillations (nodding motions) about the mean precession motion can occur, with an period $P_\mathrm{n,s}$ defined as 

\begin{equation} \label{Pnods}
P_\mathrm{n,s} = \frac{1}{2}\left(P_\mathrm{orb,s}^{-1}+P_\mathrm{prec,s}^{-1}\right)^{-1},
\end{equation}
\\where the retrograded nature of the disc precession was already considered in \autoref{Pnods} (e.g., \citealt{1997ApJ...478..527K} for further details).

The predicted values for $P_\mathrm{n,s}$ from \autoref{Pnods} are also plotted in {\autoref{fig:SMBHB_Periods_ballistic} and} \autoref{fig:SMBHB_Periods}. Considering the allowed values for the orbital period of the secondary BH, only non-ballistic precession scenario admits nodding motion of the jet, with $P_\mathrm{n,s}$ ranging roughly from 0.07 yr to 20 yr. Such periodic modulations introduce an additional oscillation about the precession motion of the disc/jet with an intrinsic amplitude, $\varphi_\mathrm{n}$, given as (e.g., \citealt{1997ApJ...478..527K})

\begin{equation} \label{varphinods}
\varphi_\mathrm{n} = \frac{P_\mathrm{n,s}}{P_\mathrm{prec,s}}\tan\varphi_0,
\end{equation}
\\resulting in $\varphi_\mathrm{n}$ roughly between 0\fdg003 to 1\fdg0 for AO\,0235+164. 

The impact of nodding oscillations on $\beta_\mathrm{obs}$ and $\eta$ can be quantified substituting $\varphi_0$ by $\varphi_0\pm\varphi_\mathrm{n}$ in \autoref{exs} and \autoref{eys}. Negligible effects on $\beta_\mathrm{obs}$ and $\eta$ are obtained considering $\varphi_\mathrm{n}\sim 0\fdg003$. However, the upper limit of about 1\fdg0 for $\varphi_\mathrm{n}$ introduces changes in $\beta_\mathrm{obs}$ ranging approximately from $-16.0c$ to $1.6c$. These extreme values occurs when the jet is closer to the line of sight ($\phi\la 4^\circ$), while amplitude changes lower than $0.4c$ are expected for $\phi\ga 15^\circ$ considering our best precession model ($P_\mathrm{prec,obs}=8.39$ years). In the case of $\eta$, oscillations between -31\degr and 31\degr around the jet position angle due to precession are expected when the jet is closer to the line of sight, reducing to less than 2\degr for $\phi\ga 15^\circ$. Interestingly, there are a bunch of jet components with $\eta\sim200^\circ$ and $\beta_\mathrm{obs}\sim 10c$ that are systematically below of the predictions from our jet precession models in the $\beta_\mathrm{obs}-\eta$ plots shown in \autoref{fig:Prec}. This particular value of position angle occurs when the jet is closer to the line of sight ($\phi\sim1\fdg4$), exactly when the nodding oscillations can reach their largest amplitudes. Indeed, a decrease of $16c$ in $\beta_\mathrm{obs}$ would be enough to reconcile our jet precession models with the apparent speeds of those components, even though a variation of 31\degr in $\eta$ is not capable of reconciling model and the mean position angles of the jet components by itself. Therefore, the imprints of the nodding motions might be present in the kinematic of the parsec-scale jet components of AO\,0235+164. Further analyses including nodding oscillations in the framework of jet precession are necessary to confirm it, as well as the high-cadence interferometric monitoring of the jet activity in this source.

\subsection{Short-term periodicities and the SMBHB in AO 0235+164}\label{subsec:Periodicity}

As mentioned anteriorly, several periodicities in the continuum light curves of AO\,0235+164 have been reported in the literature. The 8-yr periodicity at optical wavelengths may be attributed to the periodic Doppler factor changes driven by the parsec-scale jet precession happening in a clockwise sense. Moreover, the most powerful flares seen at $\gamma$-rays coincides in time with the highest values of the Doppler boosting factor in our best precession model (bottom-left panel in \autoref{fig:Prec}), reinforcing such geometrical interpretation for the 8-yr periodicity.

At radio frequencies, periodicities between 2 and 3 years are often, with an optical counterpart of 2.95 years also reported in \citet{2002A&A...381....1F}. Similar behaviour is found for periodic variabilities occurring at a time-scale of 5-6 years. It is worth to verify whether both periodicities could be accommodated simultaneously in the SMBHB scenario responsible for the jet precession in AO\,0235+164. The main idea is to link those common short-term periodicities to the secondary black hole's orbital period and the jet's nodding motion. To better quantify it, we chose the periods 5.4 and 2.8 determined by \citet{2006ApJ...650..749L}\footnote{\citet{2006ApJ...650..749L} also considered a SMBHB as a natural candidate for explaining the six quasi-periodic oscillations that they detected at radio light curves of AO\,0235+164.} as representative of such periodicities in the light curve of this source. We converted these periods to their values at the source's reference frame, indicating them by the dotted horizontal lines in the two panels of \autoref{fig:SMBHB_Periods}. Comparing them with the predicted orbital and nodding periods, tighter ranges for $q_\mathrm{BH}$ and $R_\mathrm{out,p}$ are obtained: $0.04\la q_\mathrm{BH}\la 1$ and $0.001\la R_\mathrm{out,p} \mathrm{(pc)}\la 0.007$ for $s=0$, and $0.1\la q_\mathrm{BH}\la 1$ and $0.002\la R_\mathrm{out,p} \mathrm{(pc)}\la 0.007$ for $s=-2$. Therefore, the SMBHB scenario explored in this work, together with a {\bf non-ballistic} jet precession are able to provide a physical interpretation for three periodicities in the light curves of AO\,0235+164.

\section{Alternative scenarios} \label{AlterScenarios}

In the preceding section, we examined the feasibility of a SMBHB, wherein the secondary SMBH orbits the primary in a non-coplanar plane relative to the primary accretion disc, as a potential explanation for the precession of the parsec-scale jet in AO\,0235+164. However, it is not the only mechanism capable of driving jet precession in astrophysical objects. In addition, the data scattering around the predictions from the CE-optimized jet precession models in \autoref{fig:Prec} (particularly in the $\beta_\mathrm{obs}$ versus $\eta$ plots) suggests possible additional or alternative scenarios not considered previously in this work, such as intrinsic curved jet trajectories, warped accretion disc instabilities, and Lense-Thirring precession. Here, we explore such possibilities in the context of AO\,0235+164.

\subsection{Geodetic precession} \label{subsec:GeoPrec}

This mechanism also involves the presence of a SMBHB in the core of AGNs. Originally proposed for binary pulsars in our Galaxy (e.g., \citealt{1975ApJ...199L..25B}), the geodetic precession must occur when there is a misalignment between the spin axis of the primary and/or secondary SMBH and the direction of orbital angular momentum of the binary system. If the jet is aligned with the spin axis of one of the SMBHs, it will also precesses at a period $P_\mathrm{prec}^\mathrm{geo}$ (e.g. \citealt{2019MNRAS.482..240K}):

\begin{equation}\label{eq:GeoPrec}
   \left(\frac{P_\mathrm{prec}^\mathrm{geo}}{\mathrm{Myr}}\right)\cong 124\frac{(1+q_\mathrm{BH})^2}{q_\mathrm{BH}(3q_\mathrm{BH}+4)}\left(\frac{d_\mathrm{BH}}{\mathrm{1 pc}}\right)^{5/2}\left(\frac{M_\mathrm{BH}}{\mathrm{10^9 \mathrm{M}_\odot}}\right)^{-3/2},  
\end{equation}
\\considering a circular orbit. 

In the case of AO 0235+164, an upper limit for $d_\mathrm{BH}$ can be derived from \autoref{eq:GeoPrec} by setting $q_\mathrm{BH}=1$ and adopting $M_\mathrm{BH}=10^{8.54\pm 0.12}$ M$_{\sun}$. Assuming a geodetic (non-ballistic) precession period of $P_\mathrm{prec}^\mathrm{geo}\sim 188$ years, we obtain $d_\mathrm{BH}\la 3.1$ mpc, which yields $\tau_\mathrm{GW,s}\sim 5251$ yr. This suggests that the SMBHB would remain stable against gravitational-wave losses if geodetic precession is indeed responsible for the jet precession in AO 0235+164. By contrast, for ballistic precession with $P_\mathrm{prec}^\mathrm{geo}\sim 4.32$ yr, we find $d_\mathrm{BH}\la 0.7$ mpc and $\tau_\mathrm{GW,s}\sim 13$ yr, indicating that the SMBHB would be unstable to gravitational-wave losses in this scenario.

\subsection{Warped accretion disc by the Lense-Thirring effect} \label{subsec:LTeffect}

Frame dragging produced by a Kerr black hole causes precession of a particle if its orbital plane is inclined in relation to the equatorial plane of the black hole. The precession angular velocity $\Omega_\mathrm{LT}$ due to the Lense-Thirring effect (LT) is defined as \citep{1918PhyZ...19..156L}:

\begin{equation}\label{eq:OmegaLT}
    \Omega_\mathrm{LT}(r) = \frac{2G}{c^2}\frac{J_\mathrm{BH}}{r^3},
\end{equation}
\\where $r$ is the radial distance from the spinning SMBH with an angular momentum, $J_\mathrm{BH}$,

\begin{equation}\label{eq:JBH}
    J_\mathrm{BH}=a_\ast \frac{GM_\mathrm{BH}^2}{c},
\end{equation}
\\where $a_\ast$ is a dimensionless parameter ranging from -1 to 1, defined as the ratio of the angular momentum of the compact object and that of a Kerr black hole rotating at its maximal velocity.

Besides precession of the angular momenta of the spinning SMBH and the disc around the total angular momentum of the system, the combined influence of the LT effect and the internal viscosity of the accretion disk also induces an alignment (or counteralignment in some cases; e.g. \citealt{2005MNRAS.363...49K, 2006MNRAS.368.1196L}) between the angular momenta of a Kerr SMBH and its accretion disc. This phenomenon, referred to as the Bardeen-Petterson effect (BP) \citep{1975ApJ...195L..65B}, predominantly affects the innermost regions of the disc, as the LT effect operates over a limited range due to its dependence on $r^{-3}$. In contrast, the outer regions of the accretion disc generally retain their original orientation. The boundary between these two regions, known as the Bardeen-Petterson radius, $R_\mathrm{BP}$, is primarily determined by the physical characteristics of the accretion disc (e.g. \citealt{1985MNRAS.213..435K, 1997MNRAS.285..394I, 2000MNRAS.315..570N}), with this transition occurring smoothly. 

The shape of a steady-state accretion disc under BP torques was derived analytically by \citet{1996MNRAS.282..291S} under the assumption of a disc with constant surface density and constant viscosities acting azimuthally and vertically along it. \citet{2009MNRAS.400..383M} extended the Scheuer \& Feiler's results to the case where viscosities and surface densities are power laws in the distance from the SMBH, while \citet{2009MNRAS.398.1900C} also considered power-law viscosities with indices not necessarily equal. The alignment and precession time-scales were also calculated by \citet{1996MNRAS.282..291S} and \citet{2009MNRAS.400..383M} assuming an infinite outer disc radius (or an outer radius much larger than the BP (warping) radius; see Section 5 in \citealt{2009MNRAS.400..383M} for further information). \citet{Nandi2021} and \citet{2024MNRAS.530.4902S} showed that the precession periods of their sample of radio galaxies are fully compatible with those expected from a steady-state BP accretion disc as described by \citealt{2009MNRAS.400..383M}. However, this formalism predicts much longer precession time-scales ($\ga 10^6$ years) than the jet precession period of 4.32 years (or $\sim188$ years for the non-ballistic precession) estimated for AO\,0235+164 in this work. In these calculations, we employed equations (54) and (57) from \citet{2009MNRAS.400..383M}, along with estimates for the surface density of a power-law accretion disc, following the approach of \citet{2006ApJ...653..112C}. We have also assumed a dimensionless viscosity parameter, $\alpha$ \citep{1973A&A....24..337S}, equal to 0.1, and a bolometric luminosity of $7\times10^{-3}L_\mathrm{Edd}$ \citep{2006ApJ...650..749L}, where $L_\mathrm{Edd}$ denotes the Eddington luminosity in those calculations.

Although numerical simulations have confirmed the classical configuration of a BP accretion disc (e.g. \citealt{2000MNRAS.315..570N, 2007MNRAS.381.1287L, 2010MNRAS.405.1212L, 2012MNRAS.421.1201N, 2018MNRAS.474L..81L}), this picture can be fundamentally altered by strong magnetic fields, non-linear effects arising from high inclination angles between the angular momenta of the SMBH and the disc, or for a disc in a low-viscosity regime. Considering non-linear fluid dynamics of an warped accretion disc (e.g. \citealt{1999MNRAS.304..557O, 2000MNRAS.317..607O}), \citet{2012MNRAS.421.1201N} found that when $\alpha\la 0.3$ and the initial angle of misalignment between the disc and hole is $\ga 45\degr$, frame dragging by the Kerr BH tears the disc apart into differentially precessing smaller discs. For a smaller tilted angle of about 30\degr, the disc breaking could happen for a low-viscosity regime ($\alpha<<0.2$). Numerical simulations corroborate the possibility of a disc be broken in individual small independent (precessing) discs by the Lense-Thirring effect induced by a single black hole (e.g., \citealt{2006MNRAS.368.1196L, 2012ApJ...757L..24N, 2015MNRAS.448.1526N, 2021MNRAS.507..983L, 2021ApJ...909...81R, 2021ApJ...922..243D, 2023MNRAS.518.1656M, 2023ApJ...955...72K}).

Analytical estimates for the upper limit of the radius where the accretion disc can break, $R_\mathrm{break}$, were firstly derived in a diffusive and wave regimes by \citet{2012ApJ...757L..24N} and \citet{2015MNRAS.448.1526N}, respectively. In practice, this upper limit depends on several parameters (e.g., $\alpha$, disc's aspect ratio, black hole spin, etc.), but for reasonable values a minimum initial tilt angle of the order of one degree can be obtained\citep{2012ApJ...757L..24N, 2015MNRAS.448.1526N}. The precession angle of about 9\degr found in this work (\autoref{tab:PrecParams}) is barely above this theoretical lower limit, which means the tearing of the accretion disc might occur in AO\,0235+164. 

For a disc tilted by $\sim 9\degr$, the (theoretical) breaking radius must occur at a distance smaller than about one thousand gravitational radius for $\alpha\sim 0.01 - 0.3$ and disc's aspect ratio between 0.001 and 0.1, depending on the SMBH spin. Assuming that the inner disc formed by the tearing process is precessing rigidly due to the LT effect, we can use the formalism introduced by \citet{2002ApJ...573L..23L} to check whether the jet precession period inferred in this work is compatible with this scenario. \citet{2004ApJ...616L..99C} analysed the possibility of the spin-induced precession in a small sample of AGNs and in Sgr A$^\ast$, calculating the ratio $P_\mathrm{prec,s}/M_\mathrm{BH}$ as a function of $a_\ast$. In the case of AO\,0235+164, $P_\mathrm{prec,s}/M_\mathrm{BH}\sim5.4\times10^{-7}$ yr M$_{\sun}^{-1}$ for the non-ballistic jet precession scenario or $P_\mathrm{prec,s}/M_\mathrm{BH}\sim1.2\times10^{-8}$ yr M$_{\sun}^{-1}$ for the ballistic one, implying in a conservative value for the outer radius of the inner disc smaller than about some thousands of gravitational radius, $R_\mathrm{g}(=GM_\mathrm{BH}/c^2)$ (see figure 1 in \citealt{2004ApJ...616L..99C}). It is in fair agreement with the predicted upper limit for the breaking radius in AO\,0235+164 ($\la 1000 R_g$), suggesting the feasibility of the spin-induced precession acting upon a broken accretion disc in this blazar. 

Note that the LT effect and torques generated in a SMBHB may coexist (including in AO\,0235+164), bringing much more complexity to the evolution of accretion discs  in such systems. \citet{2013MNRAS.434.1946N} analysed the case of a circumbinary accretion disc that is misaligned to the orbital plane of a SMBHB and found that discs are susceptible to tearing for almost all tilt angles. \citet{2015JCAP...07..005H} extended such analyses to the case of a SMBHB in a eccentric orbit. \citet{2014MNRAS.441.1408T} explored the dynamics of warped accretion discs around BHs, focusing on the interplay of various torques, including LT, viscous torques, companion, and self-gravitational torques. \citet{2015MNRAS.449.1251D} showed that tilted discs inside a binary system are susceptible to tearing from the outside in, because of the gravitational torque from the secondary (no LT torques are included in the calculations). From 1D numerical calculations, \citet{2020MNRAS.496.3060G} studied the non-linear dynamics of warped accretion discs under the influence of both relativistic frame dragging and binary companion. One of the results they found was the angular momentum of each BH either achieve complete alignment with the angular momentum of the disc or reach a critical obliquity, beyond which stationary solutions cease to exist, leading to a broken disc. Hydrodynamical simulations by \citet{2022MNRAS.509.5608N} corroborate those findings, also showing that when disc breaks, the ability of BHs and disc to align is compromised and in some cases even prevented as the binary separation shortens.

\subsection{Bent trajectories and dual jets in a binary system} \label{subsec:BentJet_DualJets}

Jet precession is not the only interpretation for temporal changes in the apparent speed, position angle and flux density of jet components. Helical motions of the jet plasma can also produce bent trajectories on the plane of the sky (e.g., \citealt{1995A&A...302..335S}). Orbital motions in a SMBHB (e.g., \citealt{1999A&A...347...30V}), internally rotating jet flows (e.g., \citealt{1992A&A...255...59C}), or even angular variations in the flow direction (e.g., associated with a precessing jet; \citealt{2000ApJ...533..176H}) are the usual mechanisms invoked in the literature to excite helical patterns in jets.

In the case of AO\,0235+164, helical jet motions have also been proposed to explain either periodic variabilities in its continuum spectrum \citep{2000A&A...357...84Q, Ostero, 2006A&A...459..731R, 2009ApJ...696.2170R, 2021MNRAS.501.5997T, 2023MNRAS.518.5788O} or bent trajectories of jet components \citep{2018MNRAS.475.4994K}. Indeed, jet interactions with the external medium can also change the velocity of the jet components and/or deflecting them from their initial trajectories as they recede from the core, producing curved jet paths on the plane of the sky (e.g., \citealt{1999ApJ...518L..87A, 2003ApJ...589L...9H, 2008ASPC..386..240L, 2020NatCo..11..143A}). Trailing compressions triggered by pinch-mode jet-body instabilities caused by the propagation of a strong perturbation (superluminal jet component) may also exhibit bent trajectories (e.g., \citealt{2001ApJ...549L.183, 2001ApJ...561L.161G, 2013A&A...551A..32F}).

Another scenario that could account for both the continuum variability and the kinematics of the parsec-scale jet components in AO\,0235+164 involves dual jets produced by individual black holes in a binary system. Although, to the best of our knowledge, this possibility has not previously been suggested for AO\,0235+164, it has been explored in other blazars (e.g., \citealt{2019A&A...621A..11Q, 2021A&A...653A...7Q}), also offering a potential framework in which the collision of two relativistic parsec-scale jets might generate high-energy neutrinos \citep{2019A&A...630A.103B}.

If this scenario applies to AO\,0235+164, it is conceivable that some jet components identified in this study may be associated with a secondary black hole, rather than strictly following our precession models shown in \autoref{fig:Prec}. In particular, the jet components C12–C19 (with apparent speeds of approximately 10$c$ and position angles around –200\degr) are systematically below the curves predicted by our precession models in the $\beta_\mathrm{app}-\eta$ plots, suggesting that these components might have been ejected by an active secondary black hole.

\section{Final Remarks}
\label{sec05}

In this study, we analysed public radio interferometric maps of AO\,0235+164 at 15 GHz and 43 GHz, obtained over nearly 30 years. They were modelled using two-dimensional elliptical Gaussian components, following the criteria of \citet{2014MNRAS.441..187C}. The CE global optimization technique was applied to all 203 interferometric maps studied in this work (41 maps at 15 GHz and 162 maps at 43 GHz) and was used to estimate the structural parameters of these Gaussian components. Our main results are summarized as follows.

\begin{itemize}

\item We kinematically identified 36 parsec-scale jet components moving relativistically from the source's radio core. Their apparent velocities range from 3.0$c$ to 34.3$c$, and their mean position angles are distributed across all quadrants of the sky;

\item Based on the kinematic of the jet components, we estimated the minimum jet bulk Lorentz factor to be 34.3 $\pm$ 6.6 
and the maximum jet viewing angle to be $36\fdg7 \pm 7\fdg8$; 

\item The brightness temperature of the core region at 15 and 43 GHz ranges from about $7 \times 10^{10}$ K to $2.2\times 10^{13}$ K, consistent with previous estimates and confirming that AO\,0235+164 can reach $T_\mathrm{B,rest}$ as high as $10^{13}$ K;

\item Differences in the apparent velocity and position angle of the jet components suggest that the direction of the parsec-scale jet changes over time. An analytical precession model was used to simultaneously fit the right ascension and declination offsets, mean position angles, apparent velocities of the jet components, and the brightness temperature of the core region (for the first time in the literature);

\item The seven free parameters of the jet precession model were determined using the CE global optimisation technique for both clockwise and counter-clockwise precession. The parameters $\gamma$, $\phi_0$ and $\varphi_0$ obtained for both senses agree with each other within their respective uncertainties. The jet precession period at the observer's reference frame was found to be about 8.4 years for clockwise precession, slightly longer than the 6.0-year period for counter-clockwise precession. Both precession periods agree with some of the periodicities detected in the light curves of AO\,0235+164;

\item Both jet precession models fit the observational constraints fairly well, with clockwise precession being slightly favoured based on its merit function value (see \autoref{tab:PrecParams}). The coincidence between the main gamma-ray flares seen in the public {\it Fermi} light curve of AO\,0235+164 and the peaks of the Doppler factor predicted by the clockwise jet precession provides an additional advantage over the counter-clockwise precession (see the bottom panels in \autoref{fig:Prec}). However, as $S(k_\mathrm{max})$ for clockwise and counter-clockwise precession differs by less than a factor of two, counter-clockwise precession cannot be completely ruled out for AO\,0235+164;

\item We investigated the possibility of the parsec-scale jet precession in AO\,0235+164 is driven by a secondary SMBH in a non-coplanar circular orbit around a primary SMBH, assuming that its jet comes from the primary accretion disc torqued by the secondary SMBH. We found that the orbital period of the secondary must be roughly between 0.5 and 1 year for the ballistic precession or between 0.2 and 40 years in the non-ballistic scenario, which implies a separation between the primary and secondary black holes between 0.002 and 0.004 pc and 0.001 and 0.05 pc, respectively. Even though a so compact system cannot be angularly resolved by the current instrumentation, there is some room for gravitational wave experiments like Pulsar Timing Arrays (e.g., \citealt{2023A&A...678A..48E}) might detect such SMBHB in AO\,0235+164;

\item \citet{1982ApJ...260..780K} demonstrated that short-term oscillations (nodding motions) around the mean precession motion can occur in such SMBHB. For AO\,0235+164 under a non-ballistic parsec-jet precession, the nodding period in the source's reference frame is roughly between 0.07 and 20 years, causing an amplitude oscillation around the precession motion of the disc between 0\fdg003 and 1\fdg0. Considering the later value, additional changes in the apparent velocities and position angles of the jet components between $-16c$ and $1.6c$ and $-31\degr$ and $31\degr$ are respectively expected in this case. Indeed, a decrease of $16c$ in $\beta_\mathrm{obs}$ would be sufficient to reconcile our jet precession models with the jet components having $\eta \sim 200\degr$ and $\beta_\mathrm{obs} \sim 10c$ in the $\beta_\mathrm{obs}-\eta$ plots shown in \autoref{fig:Prec};

\item We also explore the possibility that the radio/optical periodicities of 5-6 years and 2-3 years are respectively associated with the orbital period of the secondary SMBH and the jet's nodding motion, which better restricted the possible values of $q_\mathrm{BH}$ for the putative binary system in AO\,0235+164. Thus, the SMBHB scenario explored in this work, along with the jet precession, can provide a physical interpretation for three periodicities found in the light curves of AO\,0235+164;

\item A SMBHB with the secondary SMBH orbiting the primary in a non-coplanar plane relative to the primary accretion disc is not the only physical mechanism that could induce disc/jet precession. In the case of geodetic precession (e.g., \citealt{1975ApJ...199L..25B}), a separation between the secondary and primary SMBHs must be smaller than about 3 mpc to make the parsec-scale jet of AO\,0235+164 precesses at the rate inferred in this work for non-ballistic precession. A SMBHB like this is relatively stable in terms of gravitational wave losses ($\tau_\mathrm{GW,s}\sim 5251$ years). However, $d_\mathrm{BH}\la 7\times 10^{-4}$ pc ($\tau_\mathrm{GW,s}\sim 13$ years) for the ballistic jet precession scenario, casting doubts on geodetic precession mechanism in this particular case;

\item The Bardeen-Petterson effect \citep{1975ApJ...195L..65B} is a potential candidate for driving precession in misaligned accretion discs surrounding Kerr black holes. Considering the analytical formulae for the precession time-scale derived by \citet{2009MNRAS.400..383M} for a steady-state disc with an outer radius much larger than the warping radius, we obtain precession periods longer than $10^6$ years in the source's reference frame, incompatible with the precession periods found in this work for AO\,0235+164. However, analytical work and numerical simulations show that the disc may break apart in some situations, with each segment precessing independently. The inner disc precesses at a faster rate, potentially explaining the precession of the jet in AO\,0235+164. Indeed, the spin-induced precession (e.g. \citealt{2002ApJ...573L..23L, 2004ApJ...616L..99C} is able to recover the precession period inferred in this work if the outer radius of the inner disc is roughly smaller than some thousands of gravitational radius;

\item Although the kinematic of the 36 jet components identified in this work is compatible with ballistic motions, nine of them present some scatter
around their receding motions implied by the ballistic scenario. Helical motions of the jet plasma could be behind of those bent trajectories, which might be driven by the own jet precession \citep{2000ApJ...533..176H}, orbital motions in a SMBHB (e.g., \citealt{1999A&A...347...30V}), or in internally rotating jet flows (e.g., \citealt{1992A&A...255...59C});

\item An alternative explanation for the variability and kinematics of AO\,0235+164 involves dual relativistic jets produced by a binary black hole system. Though not previously proposed for this source, such scenarios have been invoked for other blazars (e.g., \citealt{2019A&A...621A..11Q, 2021A&A...653A...7Q}). In this framework, jet components C12-C19 — exhibiting apparent velocities of $\sim 10c$ and position angles of about $-200\degr$ — may originate from a secondary black hole, potentially deviating from the modelled jet precession curves. This interpretation could also imply conditions favourable for high-energy neutrino production via jet collisions \citep{2019A&A...630A.103B}.

\end{itemize}

Although our jet precession models have captured the general trends of the kinematic of the parsec-scale jet of AO\,0235+164, future studies, including high-cadence interferometric monitoring of the source's activity, are essential to confirm them, as well as the existence of a SMBHB in the nucleus of AO\,0235+164.

\section*{Acknowledgements}

F.B.S.J. thanks the Brazilian agency CAPES for financial support. We thank the anonymous referee for the constructive comments and suggestions, which helped improve the original version of this manuscript.
This research has made use of data from the MOJAVE database that is maintained by the MOJAVE team \citep{2018ApJS..234...12L}. This study makes use of 43 GHz VLBA data from the VLBA-BU Blazar Monitoring Program (VLBA-BU-BLAZAR; http://www.bu.edu/blazars/VLBAproject.html), funded by NASA through the Fermi Guest Investigator Program. The VLBA is an instrument of the National Radio Astronomy Observatory. The National Radio Astronomy Observatory is a facility of the National Science Foundation operated by Associated Universities, Inc.

\section*{Data Availability}

The data generated in this research will be shared on reasonable request to the corresponding author.


\bibliographystyle{mnras}
\bibliography{biblio} 


\bsp	
\label{lastpage}
\end{document}